\documentclass[aps,prl,reprint,twocolumn,preprintnumbers,floatfix,nofootinbib]{revtex4-1}

\usepackage[T1]{fontenc} 
\usepackage[utf8]{inputenc}

\usepackage[dvipdfm]{graphicx}
\usepackage{mathrsfs}
\usepackage{amssymb}
\usepackage{verbatim}
\usepackage{color}
\usepackage[dvipsnames]{xcolor}
\usepackage{multirow}
\usepackage{amsmath}
\usepackage{subfig}
\usepackage{slashed} 
\usepackage{float}
\usepackage[normalem]{ulem}
\usepackage{sidecap}
\usepackage{hyperref}
\usepackage{cancel}
\usepackage{enumerate}
\hypersetup{pdfstartview=FitV,colorlinks=true,linkcolor=blue,citecolor=red,filecolor=black,urlcolor=blue}

\newcommand*\diff{\mathop{}\!\mathrm{d}}

\newcommand{\beq}{\begin{equation}}
\newcommand{\eeq}{\end{equation}}
\newcommand{\bea}{\begin{eqnarray}}
\newcommand{\eea}{\end{eqnarray}}

\def\del{\partial}
\def\lrd{\overset{\leftrightarrow}{\del}}

\usepackage[parfill]{parskip}
\begin{document}
\sloppy 

\preprint{KIAS-P21021}

\vspace*{1mm}

\title{Conformal Portal to Dark Matter} 

\author{Kunio Kaneta$^{a}$}
\email{kkaneta@kias.re.kr}
\author{Pyungwon Ko$^{a}$}
\email{pko@kias.re.kr}
\author{Wan-Il Park$^{b}$}
\email{wipark@jbnu.ac.kr}

\affiliation{$^a$School of Physics, Korea Institute for Advanced Study, Seoul 02455, Korea}
\affiliation{$^b$Division of Science Education and Institute of Fusion Science, Jeonbuk National University, Jeonju 54896, Korea}

\date{\today}

\begin{abstract}
We propose a new portal coupling to dark matter by taking advantage of the nonminimally coupled portal sector to the Ricci scalar.
Such a portal sector conformally induces couplings to the trace of the energy-momentum tensor of matters including highly secluded dark matter particles.
The portal coupling is so feeble that dark matter is produced by freeze-in processes of scatterings and/or the decay of the mediator.
We consider two concrete realizations of the portal: conformally induced Higgs portal and conformally induced mediator portal.
The former case is compatible with the Higgs inflation, while the latter case can be tested by dark matter direct detection experiments.

\vskip 1cm

\vskip 1cm

\end{abstract}

\maketitle

\setcounter{equation}{0}

\section{I. Introduction}

Although the existence of the hypothetical matter, dark matter, is evident and has extensively been studied so far, its particle nature remains unidentified yet.
The paradigm of the Weakly Interacting Massive Particle (WIMP) has led investigations to uncover the dark matter for a long time \cite{Gunn:1978gr}.
The WIMP is an attractive Beyond-the-Standard-Model (BSM) particle in that its nature, such as mass scale and interaction strength, is close to that of the Standard Model particles, except that WIMP has to be neutral and stable or long-lived enough to explain the observed dark matter relic density. In modeling the WIMP, {\it portal} couplings to look into the dark matter sector from the visible sector is considered as a useful concept, and a number of portals have been considered so far, including Higgs portal \cite{Silveira:1985rk,McDonald:1993ex,Burgess:2000yq,Davoudiasl:2004be,Mambrini:2011ik,Baek:2011aa,Djouadi:2011aa,Lebedev:2011iq,Djouadi:2012zc,Baek:2012se,Baek:2012uj,Han:2015hda,Casas:2017jjg}, Z portal \cite{Arcadi:2014lta,Escudero:2016gzx,Kearney:2016rng,Balazs:2017ple}, and all other BSM related portals \cite{Bai:2009ms,Baek:2013qwa,Alves:2013tqa,Arcadi:2013qia,Lebedev:2014bba,Chowdhury:2018tzw,Fuks:2020tam}.
The relic WIMP density is solely determined by the annihilation rate of the WIMPs into thermal particles by assuming that the WIMP is in thermal equilibrium at early times, rendering ultraviolet (UV) conditions of the Universe irrelevant.
The WIMP interaction strength is on the other hand as large as being accessible by using scatterings of the WIMP with nuclei, making direct detection possible.
However, direct dark matter detection experiments, such as LUX \cite{Akerib:2016vxi}, PandaX \cite{Cui:2017nnn}, and XENON1T \cite{Aprile:2018dbl}, have observed no signals so far, putting stringent constraints on the parameter space of the WIMP scenario.

Feebly Interacting Massive Particle (FIMP) has been proposed as another type of dark matter candidate \cite{Hall:2009bx}, which sheds light on new parameter spaces of theories and phenomenology of dark matter (see \cite{Bernal:2017kxu} for a review).
Similar to the gravitino dark matter \cite{Nanopoulos:1983up, Khlopov:1984pf}, the FIMP is so feebly interacting with thermal particles that it is never thermalized, which can be realized by supposing that the FIMP sector is supposed to be highly secluded from the Standard Model.
For the relic FIMP density, instead of following the equilibrium number density like the WIMP, the FIMP is produced, for instance, from the annihilation of the thermal particles, which is one-way reactions as the FIMP number density is too little to react inversely.
Such highly secluded dark matter sector is realized in many models that has super massive mediators \cite{Mambrini:2013iaa,Nagata:2015dma,Nagata:2016knk,Mambrini:2015vna,Berlin:2016vnh,Berlin:2016gtr,Heikinheimo:2016yds,Benakli:2017whb,Dudas:2017rpa,Dudas:2017kfz,Dudas:2018npp,Bernal:2018qlk,Bhattacharyya:2018evo,Banerjee:2019asa,Kang:2020afi,Bernal:2020fvw,Heurtier:2019eou} or tiny couplings \cite{Kaneta:2016vkq,Kaneta:2016wvf,Barman:2020ifq,Ema:2016hlw,Ema:2018ucl,Mambrini:2021zpp,Hashiba:2018tbu,Ahmed:2020fhc,Chianese:2020khl,Chianese:2020khl,Anastasopoulos:2020gbu,Barman:2021ugy}.  Depending on the type of relevant interactions, the FIMP is dominantly produced in either infrared or UV regime. 
For instance, if the production takes place through renormalizable couplings, the resulting FIMP abundance is insensitive to the UV, whereas it may depend on the UV when the production happens through higher dimensional operators.

In the FIMP models, although gravity is the weakest force in nature, the gravity sector often has important effects on the FIMP productions.
Besides the direct contributions  \cite{Ema:2016hlw,Ema:2018ucl,Hashiba:2018tbu,Ahmed:2020fhc,Chianese:2020yjo,Chianese:2020khl,Mambrini:2021zpp,Barman:2021ugy,Li:2021fao}, dark matter itself can arise from the space-time metric \cite{Anastasopoulos:2020gbu,Brax:2020gqg,Brax:2021gpe}.

In this work, we propose a new portal to the FIMP dark matter, where the portal coupling is induced via the conformal factor of the space-time metric and takes advantage of having a particle which has a conformally induced universal coupling to all the other particles via the trace of the energy-momentum tensor.
In particular, we show that the nonminimal coupling of the Standard Model Higgs to the Ricci scalar automatically induces a feeble coupling to dark matter and explains the observed dark matter density.
We also consider the case where a conformally induced scalar plays as a light/heavy mediator to produce dark matter, in which some scenario can be tested by direct detection experiments.

The paper is organized as follows.
In section II, we define the model for both the dark matter and conformally induced portal sectors, and discuss the dark matter productions with generic reaction rates in section III.
In sections IV and V, we compute the dark matter number density produced via the conformally induced Higgs portal and light/heavy mediator, respectively.
The phenomenology of dark matter and mediator is discussed in section VI, before concluding in section VII.

\section{II.The Model}

Our setup consists of three sectors: the Standard Model sector, the dark matter sector, and the portal sector.
To highlight the role of the portal sector, we suppose that the dark matter sector does not have any direct interactions to the Standard Model sector.
In the following, first we consider a generic framework for the portal sector, and then give a specific setup for the dark matter sector.
Throughout the paper, we use the mostly-minus convention for the metric, $+---$.

\subsection{II-A. The Portal Sector}

The portal sector is introduced as a conformal factor of the metric, $g_{\mu\nu}=C(\phi)\tilde g_{\mu\nu}$, where $g_{\mu\nu}$ defines the Einstein frame in which the gravity sector takes the canonical form with the Einstein-Hilbert action, and $\tilde g_{\mu\nu}$ defines the Jordan frame in which the gravity sector is parametrized by
\bea
S_{\rm grav}=\frac{M_P^2}{2}\int \diff^4x\sqrt{-\tilde g}C(\phi)\tilde R.
\eea
Here, $M_P$ is the reduced Planck scale ($M_P\simeq2.4\times10^{18}$ GeV), and $C(\phi)$ is an arbitrary function of a scalar field $\phi$.
It becomes clear shortly that a generic scalar field $\phi$ is taken as either the Standard Model Higgs or an additional scalar which plays a role of a mediator connecting between the Standard Model and the dark matter sector.
The conformal factor $C(\phi)$ is assumed to be expanded as
\bea
C(\phi)\simeq 1+\delta C + \cdots.
\eea
As discussed in detail in Appendix, the interactions among $\phi$, the Standard Model, and the dark matter sectors are given in the form of
\bea
{\cal L}_{\rm portal}=\frac{1}{2}\delta C g^{\mu\nu}T_{\mu\nu},
\label{eq:Lint1}
\eea
in the Einstein frame, where $T_{\mu\nu}$ contains the energy-momentum tensors for all matter fields including $\phi$.
The explicit definition and the detail in deriving the theory in the Einstein frame are given in the Appendix, where the energy-momentum tensor is introduced as a response to the variation of the metric.
\footnote{
Given the existence of $C(\phi)\tilde R$ coupling at tree level, there also be loop-level contributions that may enhance additional terms, such as $\tilde R^2$ for instance \cite{Salvio:2015kka,Calmet:2016fsr}.
Such radiative corrections involving gravity may give rise to additional portal channels whose comprehensive study is however beyond the scope of the paper.
}

\subsection{II-B. The Dark Matter Sector}

The dark matter sector is supposed to be secluded from the Standard Model sector, namely, no direct interactions between those two.
To be specific, we consider three different scenarios where dark matter is a real scalar $X$, a Dirac fermion $\chi$, or an Abelian vector $V_\mu$, whose Lagrangian is respectively given by:
\bea
&&
{\cal L}_X = \frac{1}{2}\del_\mu X\del^\mu X-\frac{1}{2}m_X^2X^2,
\\&&
{\cal L}_\chi = \overline\chi(i\cancel{\del}-m_\chi)\chi,
\\&&
{\cal L}_V = -\frac{1}{4}V^{\mu\nu}V_{\mu\nu}
+\frac{1}{2}m_V^2V_\mu V^\mu,
\label{eq:L_V}
\eea
where $V_{\mu\nu}\equiv\del_\mu V_\nu-\del_\nu V_\mu$.

For the vectorial dark matter case, the mass is put in by hand, or by St\"{u}ckelberg mechanism. In either case, perturbative unitarity in some scattering channels will be violated at high energy (or high enough temperature) and one has to consider UV completions. The simplest would be an Abelian Higgs mechanism in the dark sector \cite{Baek:2012se}, for which Eq. (\ref{eq:L_V}) is replaced by 
\bea
{\cal L}_V & = & - \frac{1}{4} V_{\mu\nu}V^{\mu\nu} + \left( D_\mu \varphi \right)^\dag D^\mu \varphi 
- \frac{\lambda_\varphi}{4\!} 
\left( \varphi^\dagger \varphi - \frac{v_\varphi^2}{2} \right)^2 
\nonumber \\
& + & 
\frac{\lambda_{\varphi H}}{4} \left( \varphi^\dagger \varphi - \frac{v_\varphi^2}{2} \right) \left( H^\dagger H - \frac{v_H^2}{2} \right)
\eea
where the dark Higgs $\varphi$ carries $U(1)_D$ charge equal to $q_\varphi$.

Dark matter phenomenology in the UV completions can be different from the effective  field theory approach. 
Nice examples are the invisible Higgs decay rate into a pair of vectorial dark matters, $\Gamma_{\rm inv} (H\rightarrow VV)$, and its correlation with spin-independent dark matter-nuclear cross section \cite{Baek:2014jga}. In particular the behavior of $\Gamma_{\rm inv} (H\rightarrow VV)$  is completely different in the limit $m_V \rightarrow 0$. 
% Also, restoration of unitarity makes the high energy behavior of production cross sections for the vectorial dark matter at high energy colliders such as ILC \cite{Ko:2016xwd,Kamon:2017yfx} and LHC/100 TeV $pp$ collider \cite{Dutta:2017sod}.
Another example is the restoration of unitarity in the high energy behavior of production cross sections for the vectorial dark matter at high energy colliders such as ILC \cite{Ko:2016xwd,Kamon:2017yfx} and LHC/100 TeV $pp$ collider \cite{Dutta:2017sod}.
Without including the dark Higgs boson, the production rate will be too much overestimated not only at high energy colliders but also at high temperature era in the early Universe.
Within this paper, however, we highlight only the conformal portal to the dark sector, and thus we will take the dark Higgs mass above the inflationary/cutoff scale so that it is never be produced after the end of inflation.
\footnote{More precisely, there are two types of cutoff scales in our framework for the vectorial dark matter. One is induced through $C(\phi)$, which is the scale where the contributions from higher dimensional operators in Eq. (\ref{eq:Lint1}) blow up. The other is induced by the breaking of $U(1)_D$, which we assume is above the other cutoff scale or the maximal temperature in the Universe.}
After integrating out the dark Higgs (having a nonzero vacuum expectation value), we obtain Eq.~(\ref{eq:L_V}) as an effective theory where the only $V_\mu$ is left as a light degree of freedom in the dark sector.
In doing so, we implicitly assume a hierarchy between the dark Higgs mass ($\sim\sqrt{\lambda_\varphi} v_\varphi$) and the vectorial dark matter mass ($\sim g_D v_\varphi$ with $g_D$ the gauge coupling of the $U(1)_D$), namely, $\sqrt{\lambda_\varphi}\gg g_D$.

The energy-momentum tensor for each dark matter case is obtained as
\bea
&&
T^X_{\mu\nu}=
\frac{1}{2}(\del_\mu X\del_\nu X +\del_\nu X\del_\mu X)-g_{\mu\nu}{\cal L}_X,
\\&&
T^\chi_{\mu\nu}=
\frac{i}{4}\overline\chi(\gamma_\mu\lrd_\nu+\gamma_\nu\lrd_\mu)\chi-g_{\mu\nu}{\cal L}_\chi,
\\&&
T^V_{\mu\nu}=
-g^{\alpha\beta}V_{\mu\alpha}V_{\nu\beta}-g_{\mu\nu}{\cal L}_V,
\eea
where $\overline\chi\lrd_\mu\chi\equiv\overline\chi(\del_\mu\chi)-(\del_\mu\overline\chi) \chi$.
From Eq.~(\ref{eq:Lint1}) it is evident that $C(\phi)$ plays the role of the portal connecting the Standard Model and the dark matter sector where the trace of the energy-momentum tensor is given by
\bea
&&
g^{\mu\nu}T^X_{\mu\nu}\simeq -(\del^\mu X)(\del_\mu X)
\label{eq:T-scalar} \\&&
g^{\mu\nu}T^\chi_{\mu\nu} = m_\chi\overline\chi\chi,
\label{eq:T-fermion} \\&& 
g^{\mu\nu}T^V_{\mu\nu} = -2m_V^2V^\mu V_\mu.
\label{eq:T-vector}
\eea
Here, for $X$ the mass term has been neglected, and for $\chi$ we have used the equation of motion.
In the following we consider two possible choices of $C(\phi)$:
\begin{enumerate}
    \item[(a)] $C(\phi=H)=1-\xi |H|^2/M_P^2$,
    \item[(b)] $C(\phi)=e^{-\phi/f}$,
\end{enumerate}
where $H$ is the Standard Model Higgs doublet, and thus (a) turns out to be a conformally induced Higgs portal scenario. In the case of (b) we consider $\phi$ as a real scalar mediator connecting between the Standard Model and the dark matter sector.

\section{III. Dark Matter Production via scattering}

The interactions between the Standard Model particles and the dark matter are feeble, so the dark matter never thermalizes.
In the following, we compute the dark matter number density produced by the annihilation of the thermal particles.

The reaction rate for the $12\to34$ process whose amplitude is written as ${\cal M}$ is given by
\begin{widetext}
\beq
R(T)=\frac{S}{1024\pi^6}\int f_1(E_1)f_2(E_2)E_1E_2\diff E_1\diff E_2\diff\cos\theta_{12}\int|{\cal M}|^2\diff\Omega_{13},
\label{eq:ReactionRate}
\eeq
\end{widetext}
where $S$ represents the symmetry factor ($S=1/2$ when the particles 3 and 4 are identical). The particles 1 and 2 are assumed to be the Higgs or $\phi$ in the most cases in the following, and therefore the distribution functions $f_1$ and $f_2$ are taken as the Bose-Einstein distribution function.
More detail is explained in Appendix of Ref. \cite{Anastasopoulos:2020gbu}.

When the spin averaged squared amplitude is written in the form of
\bea
|{\cal M}|^2=\frac{s^k}{\Lambda^{2k}}
\eea
with an integer $k$ and a suppression scale $\Lambda$,\footnote{Notice that if the reaction under consideration is not a 2 to 2 process, the squared amplitude may not be written in the dimensionless form. In such cases, one should directly compute the reaction rate by Eq.~(\ref{eq:ReactionRate}).} the reaction rate is readily computed as
\bea
R(T)= \frac{[2^k\Gamma(k+2)\zeta(k+2)T^{k+2}]^2}{128\pi^5(k+1)\Lambda^{2k}},
\eea
where we assume that the annihilating particles in the thermal bath are bosonic, while if they are fermion, a factor of $(1-2^{-(k+1)})^2$ should be multiplied \cite{Brax:2020gqg}.

Once the reaction rate is computed, the dark matter number density can be obtained by solving the Boltzmann and Friedmann equations:
\bea
&&
\frac{\diff\rho_\Phi}{\diff t}+3H\rho_\Phi=-\Gamma_\Phi\rho_\Phi,
\label{eq:BoltzmannInf}
\\&&
\frac{\diff\rho_R}{\diff t}+4H\rho_R=+\Gamma_\Phi\rho_\Phi,
\label{eq:BoltzmannRad}
\\&&
\frac{\diff n_{\rm DM}}{\diff t}+3Hn_{\rm DM}=R(t),
\label{eq:BoltzmannDM1}
\\&&
3M_P^2H^2=\rho_\Phi+\rho_R,
\label{eq:Friedmann}
\eea
where $\rho_\Phi$ and $\rho_R$ are the energy density of the inflaton $\Phi$ and radiation, respectively.
The total inflaton decay width is parametrized by $\Gamma_\Phi$.
The dark matter number density is represented by $n_{\rm DM}$.
Here, we assume the instantaneous thermalization of the inflaton decay products.\footnote{Non-instantaneous reheating and/or thermalization has been discussed in Refs. \cite{Barman:2021tgt,Bernal:2019mhf,Bernal:2020bfj,DiMarco:2018bnw,DiMarco:2019czi,DiMarco:2021xzk,Elahi:2014fsa,Garcia:2017tuj,Garcia:2018wtq,Harigaya:2014waa,Harigaya:2019tzu}.}

During the reheating epoch, the scale factor $a$ is proportional to $a\propto T^{-8/3}$, and therefore, it is convenient to rewrite Eq.~(\ref{eq:BoltzmannDM1}) as follows\footnote{Note that we assume the absence of the radiation energy density before the end of inflation.}:
\beq
\frac{\diff}{\diff T}\frac{n_{\rm DM}}{T^8}=-\frac{8}{3}\frac{R(T)}{HT^9},
\label{eq:BoltzmannDM2}
\eeq
where $H\simeq(2/5)\Gamma_\Phi(T/T_{\rm reh})^4$, the reheating temperature is defined by $\rho_\Phi(T_{\rm reh})=\rho_R(T_{\rm reh})$ leading to $\Gamma_\Phi=(5/2)\alpha_{\rm reh}T_{\rm reh}^2/M_P$ with $\alpha_i\equiv\sqrt{g_*(T_i)\pi^2/90}$, and $g_*$ is the effective degrees of freedom of $\rho_R$ \cite{Kaneta:2019zgw}\footnote{Here, we have assumed that the inflaton oscillation during the reheating is harmonic due to a quadratic potential. For anharmonic inflaton oscillation cases, see \cite{Garcia:2020eof,Garcia:2020wiy}.}.

During the radiation domination epoch where $T<T_{\rm reh}$, Eq.~(\ref{eq:BoltzmannDM1}) takes the familiar form of
\beq
\frac{\diff Y_{\rm DM}}{\diff T}=-\frac{R(T)}{HT^4},
\label{eq:BoltzmannDM3}
\eeq
where $Y_{\rm DM}\equiv n_{\rm DM}/T^3$, and $H=\alpha T^2/M_P$ with $\alpha(T)\equiv\sqrt{g_*(T)\pi^2/90}$.

For a later convenience, we parametrize the reaction rate by
\beq
R(T)\equiv\beta\frac{T^{n+6}}{\Lambda^{n+2}}
\eeq
where $\beta$ is a dimensionless coefficient.
The resultant dark matter number density depends on $n$, namely, the cases where $n<6$, $n=6$, or $n>6$.
In our case, $n$ always satisfies $n<6$, and thus we restrict ourselves to this case.
From Eq.~(\ref{eq:BoltzmannDM2}), we obtain
\beq
\frac{n_{\rm DM}(T_{\rm reh})}{T_{\rm reh}^3}=\frac{8}{3(6-n)}\frac{\beta M_PT_{\rm reh}^{n+1}}{\alpha_{\rm reh}\Lambda^{n+2}}\equiv Y_{\rm DM}(T_{\rm reh}),
\label{eq:YDMscat_Treh}
\eeq
where we take $n_{\rm DM}(T_{\rm max})=0$ with $T_{\rm max}$ ($>T_{\rm reh}$) being the maximal temperature.
With $Y_{\rm DM}(T_{\rm reh})$, from Eq.~(\ref{eq:BoltzmannDM3}), $Y_{\rm DM}$ today is computed as
\beq
Y_{\rm DM}=\frac{26+5n}{3(n+1)(6-n)}\frac{\beta M_P T_{\rm reh}^{n+1}}{\alpha_{\rm reh}\Lambda^{n+2}},
\label{eq:YDMscat}
\eeq
This expression is valid only for $n>-1$.
Notice also that Eqs.~(\ref{eq:YDMscat_Treh}) and (\ref{eq:YDMscat}) rely on the fact that the reaction rate can be parametrized by a polynomial of $T$, and thus cannot be applied to other cases, such as involving resonances.
By taking the dilution factor into account, the dark matter relic density can be obtained by
\beq
\Omega_{\rm DM}h^2\simeq 1.6\times10^8 Y_{\rm DM}\left(\frac{g_0}{g_*}\right)\frac{m_{\rm DM}}{\rm GeV},
\label{eq:Oh2scat}
\eeq
where $g_0=g_{*S}(T_0)=3.91$, and $g_*=g_{*S}(T_*)$ will be taken at a temperature $T_*$ where dark matter is dominantly produced.

\section{IV. Conformally Induced Higgs Portal}

\begin{figure*}[ht]
    \centering
    \includegraphics[width=5.2cm]{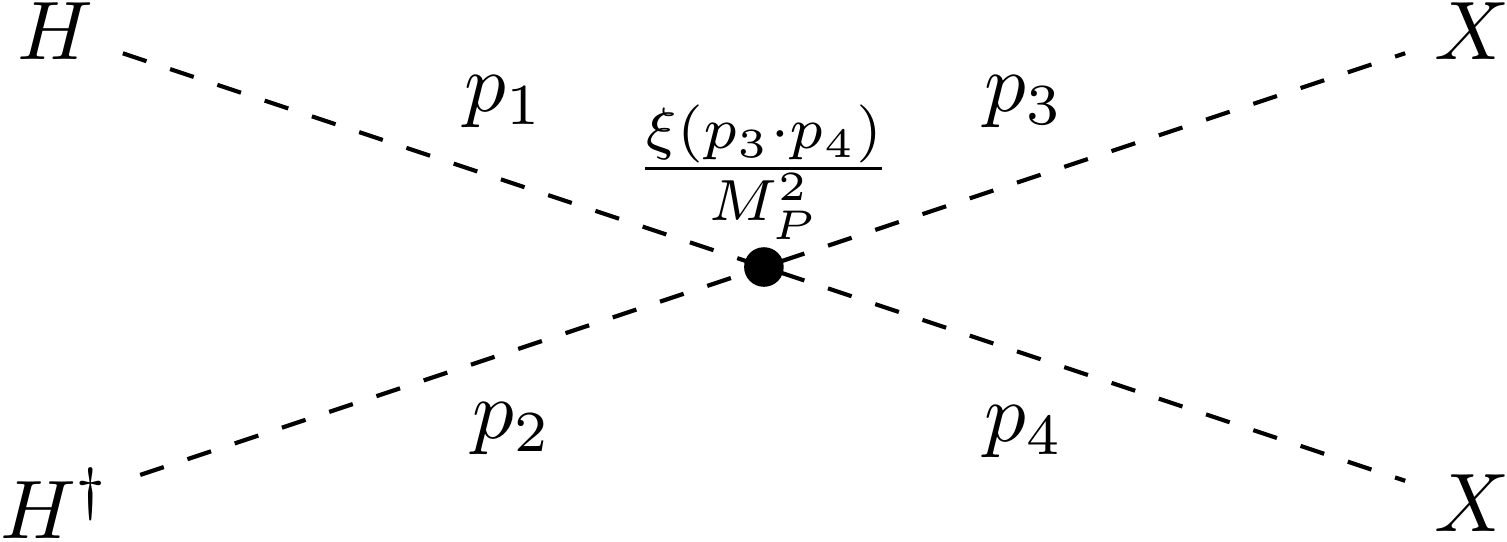}~~~
    \includegraphics[width=5.2cm]{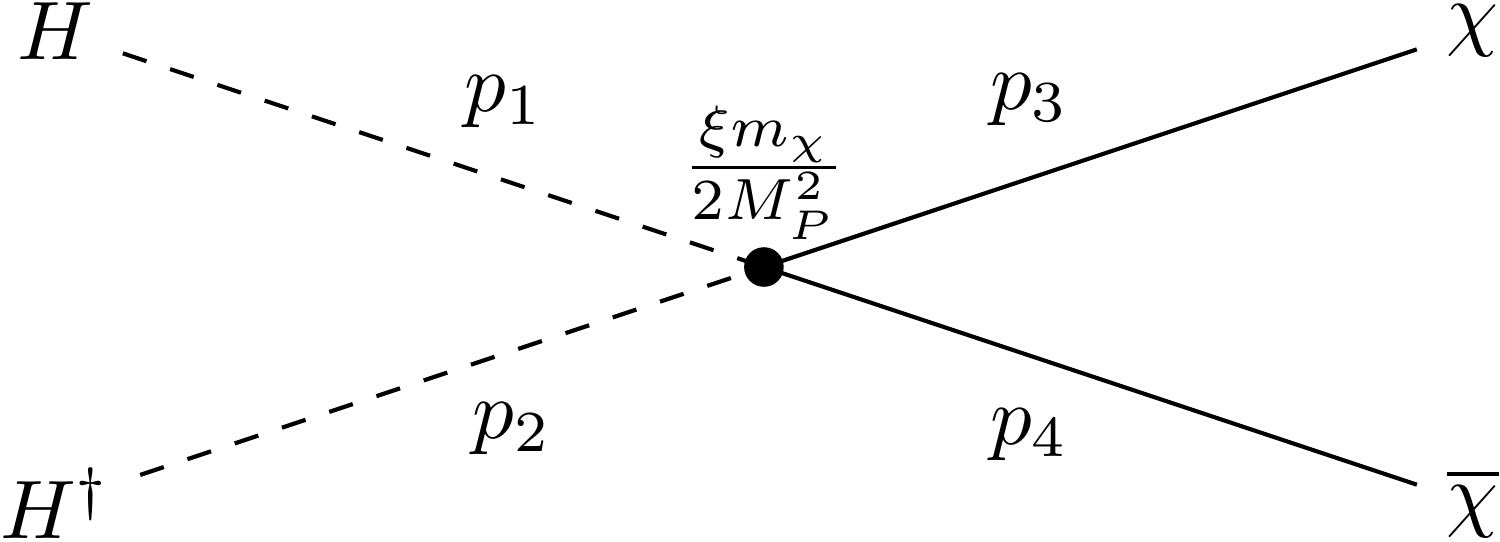}~~~
    \includegraphics[width=5.2cm]{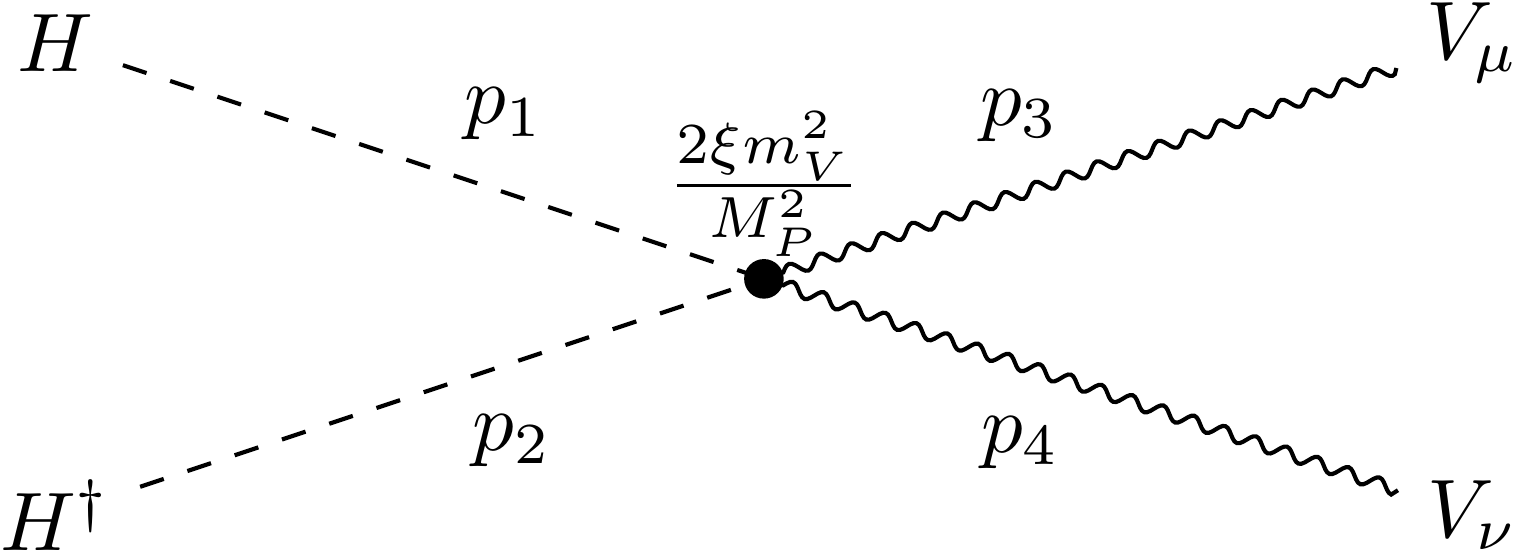}
    \caption{Feynman diagrams for the dark matter production in the conformal Higgs portal scenario.}
    \label{fig:HiggsPortal}
\end{figure*}

From Eq.~(\ref{eq:Lint1}) and the choice of $\delta C=-\xi|H|^2/M_P^2$, the portal coupling is given by
\bea
{\cal L}_{\rm portal} = -\frac{\xi}{2M_P^2}|H|^2g^{\mu\nu}T^i_{\mu\nu},
\\\nonumber
\eea
where $i=X,\chi,V$.
The relevant diagrams of the dark matter production for each type of dark matter are shown in Fig.~\ref{fig:HiggsPortal}.

The scalar dark matter is produced via $HH^\dagger\to XX$, whose reaction rate is given by
\beq
R(T)=\beta_0\frac{T^8}{M_P^4},
\label{eq:R0_HPortal}
\eeq
where $\beta_0=\xi^2\pi^3/{10800}\simeq 2.9\times10^{-3}\xi^2$.

In the same way as the scalar dark matter, the reaction rate of $HH^\dagger\to\chi\overline\chi$ is computed as
\beq
R(T)=\beta_{1/2}\frac{m_\chi^2T^6}{M_P^4},
\eeq
where $\beta_{1/2}=\xi^2\zeta^2(3)/16\pi^5\simeq3\times10^{-4}\xi^2$.
Notice that in contrast to the scalar dark matter case, the reaction rate is suppressed by the dark matter mass because of the angular momentum conservation, which requires the dark matter relatively massive to explain the observed dark matter density.

For the massive vector dark matter, it is convenient to consider the transverse and the longitudinal modes separately.
The production rate of the transverse polarization mode is given by
\beq
R(T)=\beta_1^T\left(\frac{m_V}{M_P}\right)^4T^4,
\eeq
where $\beta_1^T=\xi^2/{576}\pi\simeq 5.5\times10^{-4}\xi^2$,
leading to the yield value
\beq
Y_V^T\simeq\frac{\beta_1^T}{\alpha(m_V)}\left(\frac{m_V}{M_P}\right)^3,
\eeq
where we have assumed $m_V<T_{\rm reh}$.

On the other hand, the production of the longitudinal mode is readily computed in the same manner as the scalar dark matter case, resulting in the same expression given in Eq.~(\ref{eq:R0_HPortal}), and thus the yield value becomes
\beq
Y_V^L\simeq \frac{\beta_1^L}{\alpha_{\rm reh}}\left(\frac{T_{\rm reh}}{M_P}\right)^3,
\eeq
where $\beta_1^L=\beta_0$.
Therefore, the number density of the transverse mode becomes comparable to that of the longitudinal mode only when $m_V\sim T_{\rm reh}$, otherwise the longitudinal mode is the dominant component in the dark matter relic density, which coincides with the number density of the scalar dark matter.

With the reaction rates computed for each type of dark matter, we obtain the dark matter relic density from Eqs. (\ref{eq:YDMscat}) and (\ref{eq:Oh2scat}) as
\begin{widetext}
\bea
\frac{\Omega_{\rm DM}^{\rm scat}h^2}{0.1}\simeq \left(\frac{\xi}{10^5}\right)^2 
\left(\frac{106.75}{g_{\rm reh}}\right)^{3/2}
\left(\frac{T_{\rm reh}}{10^{11}~{\rm GeV}}\right)^3
\times
\begin{cases}
\left(\frac{m_{X,V}}{27~{\rm PeV}}\right)& \text{ (scalar/vector dark matter)} \\[10pt]
\left(\frac{m_\chi}{12~{\rm EeV}}\right)^3& \text{ (fermionic dark matter)}
\end{cases},
\eea
\end{widetext}
where $g_i\equiv g_*(T_i)$.

\subsection{IV-D. Implication of Inflation}

So far we have not specified the inflationary sector, whereas supposing a concrete inflation model gives additional information on the dark matter number density.
For instance, if we consider the Higgs inflation scenario, the nonminimal coupling is required to be $\xi\simeq49000\sqrt{\lambda}$ and $T_{\rm reh}\simeq (2\lambda/g_{\rm reh}\pi^2)^{1/4}M_P/\sqrt{\xi}\simeq2.3\times10^{15}$ GeV \cite{Bezrukov:2007ep}, where $\lambda$ is the Higgs quartic coupling and is taken as $\lambda\simeq 0.13$.
For the scalar (and vectorial) dark matter case, we thus obtain $m_X\simeq 71$ keV, while the fermionic dark matter is predicted as $m_\chi\simeq 1.7$ PeV.
Note that, as we will discuss shortly, despite a large $\xi$ being required for successful inflation, the scale where perturbative unitarity is violated is above $T_{\rm reh}$, and dark matter is produced dominantly at $T \lesssim T_{\rm reh}$ in our scenario.
Thus the perturbative computation of the dark matter production remains valid.

For a more generic inflation model, the dark matter production via the portal coupling implies an upper bound on the maximal temperature ($T_{\rm max})$, so that the cutoff scale involved should be above these temperatures.
To be more concrete, we consider a case that during the inflaton oscillation epoch the inflaton potential can be approximated by $V(\Phi)\simeq(1/2)m_\Phi^2 \Phi^2$.
As discussed in detail in Refs.~\cite{Garcia:2020eof,Garcia:2020wiy}, by solving Eqs. (\ref{eq:BoltzmannInf}), (\ref{eq:BoltzmannRad}), and (\ref{eq:Friedmann}), we obtain
\bea
T_{\rm max}=
&&
8.8\times10^{14}~{\rm GeV}\times y^{1/2}\left(\frac{106.75}{g_*(T_{\rm max})}\right)^{1/4}\nonumber
\\&&
\times\left(\frac{m_\Phi}{3\times10^{13}~{\rm GeV}}\right)^{1/4}\left(\frac{\rho_{\rm end}}{0.175m_\Phi^2M_P^2}\right)^{1/8},\label{eq:Tmax}\\
T_{\rm reh}=
&&
5.8\times10^{14}~{\rm GeV}\times y\left(\frac{106.75}{g_{\rm reh}}\right)^{1/4}\nonumber
\\&&
\times\left(\frac{m_\Phi}{3\times10^{13}~{\rm GeV}}\right)^{1/2},
\label{eq:Treh}
\eea
where we define $\Gamma_\Phi=(y^2/8\pi)m_\Phi$, and $\rho_{\rm end}$ is the inflaton energy density at the end of the inflation.
For instance, when the inflaton potential is given by the Starobinsky-like potential \cite{Starobinsky:1980te}, we obtain $\rho_{\rm end}=0.175m_\Phi^2M_P^2$ with $m_\Phi\simeq3\times10^{13}~{\rm GeV}$ \cite{Ellis:2015pla} when the end of the inflation is defined by $\diff^2 a/\diff t^2=0$ for the scale factor $a$.
By defining $\Lambda$ as a cutoff scale involved in the dark matter production, we impose $T_{\rm max}\lesssim \Lambda$ to ensure the unitarity, which results in
\begin{widetext}
\bea
\left(\frac{y}{1.3\times10^{-2}}\right)^{1/2}
&\lesssim&
\left(\frac{\Lambda}{10^{14}~{\rm GeV}}\right)\left(\frac{106.75}{g_*(T_{\rm max})}\right)^{-1/4}
\left(\frac{m_\Phi}{3\times10^{13}~{\rm GeV}}\right)^{-1/4}\left(\frac{\rho_{\rm end}}{0.175m_\Phi^2M_P^2}\right)^{-1/8}.
\label{eq:y}
\eea
\end{widetext}
In other words, once $y$ is given within the limit of Eq.~(\ref{eq:y}), the condition $T_{\rm reh}<\Lambda$ is automatically satisfied.
For instance, in the present case, the cutoff scale is supposed to be $\Lambda\sim M_P/\sqrt{\xi}=7.6\times10^{15}(10^5/\xi)^{1/2}~{\rm GeV}$, and thus the maximal and reheating temperatures needed to produce the right amount of the dark matter number density can be well below the cutoff scale.

Indeed, for the scalar and vectorial dark matter, the scale in which perturbative unitarity is violated is $\sim M_P/\sqrt{\xi}$, and for the fermionic dark matter, the scale is $M_P^2/\xi m_\chi$.
For a more detailed discussion, see Appendix.

\section{V. Conformally induced mediator}

Next, we consider the case where the conformally induced scalar $\phi$ plays a role of a mediator: $C(\phi)=e^{-\phi/f}\simeq 1-\frac{\phi}{f}+\frac{\phi^2}{2f^2}-\cdots$.
Notice that the coupling of $\phi$ looks similar to that of the dilaton which appears as a Nambu-Goldstone boson of spontaneously broken scale/conformal symmetry.
If the scale invariance is imposed to the theory, the trace of the energy-momentum tensor is proportional to a mass scale related to the violation of the scale symmetry \cite{Callan:1970ze,Coleman:1970je}.
In our case, however, we do not impose the scale invariance, and thus, $\phi$ couples to the trace of the energy-momentum tensor which is not necessarily proportional to the violation of the scale invariance, e.g., particle masses.
For more detail, see, for instance, Ref.~\cite{Nakayama:2013is}.
The relevant interactions are thus given by
\beq
{\cal L}_{\rm portal}=-\frac{1}{2}\left(\frac{\phi}{f}-\frac{\phi^2}{2f^2}+\cdots\right)g^{\mu\nu}(T^H_{\mu\nu}+T^i_{\mu\nu})
\eeq
with $i=X,\chi,V$, where the energy-momentum tensor for $H$ is given by
\bea
T^H_{\mu\nu}=
&&
2(D_\mu H)^\dagger(D_\nu H)
\nonumber\\&&
-g_{\mu\nu}[(D^\alpha H)^\dagger(D_\alpha H)+{\cal L}_{\rm Yuk}],\\
{\cal L}_{\rm Yuk}\supset
&&
-y_t\overline Q_L \widetilde H t_R + {\rm h.c.},
\eea
with omitting the Higgs potential, and $D_\mu$ being the covariant derivative.
Notice that the derivative term of the Higgs plays the essential role since the relevant processes for the dark matter production take place at high temperatures where the derivative coupling is allowed to acquire high energies and hence enhances the reaction rates.
For the derivative term in the trace of $T^H_{\mu\nu}$, we write
\beq
g^{\mu\nu}T^H_{\mu\nu}\supset-\sum_i^4(\del^\alpha h_i)(\del_\alpha h_i),
\eeq
where $h_i$ represent the real degrees of freedom of $H$, and we have dropped the gauge interaction terms in $D_\mu$.

Actual processes for the dark matter production depend on whether $\phi$ is in the thermal bath and whether $\phi$ is lighter or heavier than the dark matter.
We summarize those cases in Tab. \ref{tab:PhiPortal}.

\begin{table}[h]
\begin{tabular}{c|c|c}
     & thermal $\phi$ & nonthermal $\phi$ \\\hline
    $m_\phi<2m_{\rm DM}$ & light-thermal & light-nonthermal \\
    $m_\phi>2m_{\rm DM}$ & heavy-thermal & heavy-nonthermal
\end{tabular}
\caption{Summary of four possible cases for the dark matter production.}
\label{tab:PhiPortal}
\end{table}

\begin{figure*}[ht]
    \centering
    \includegraphics[width=5.2cm]{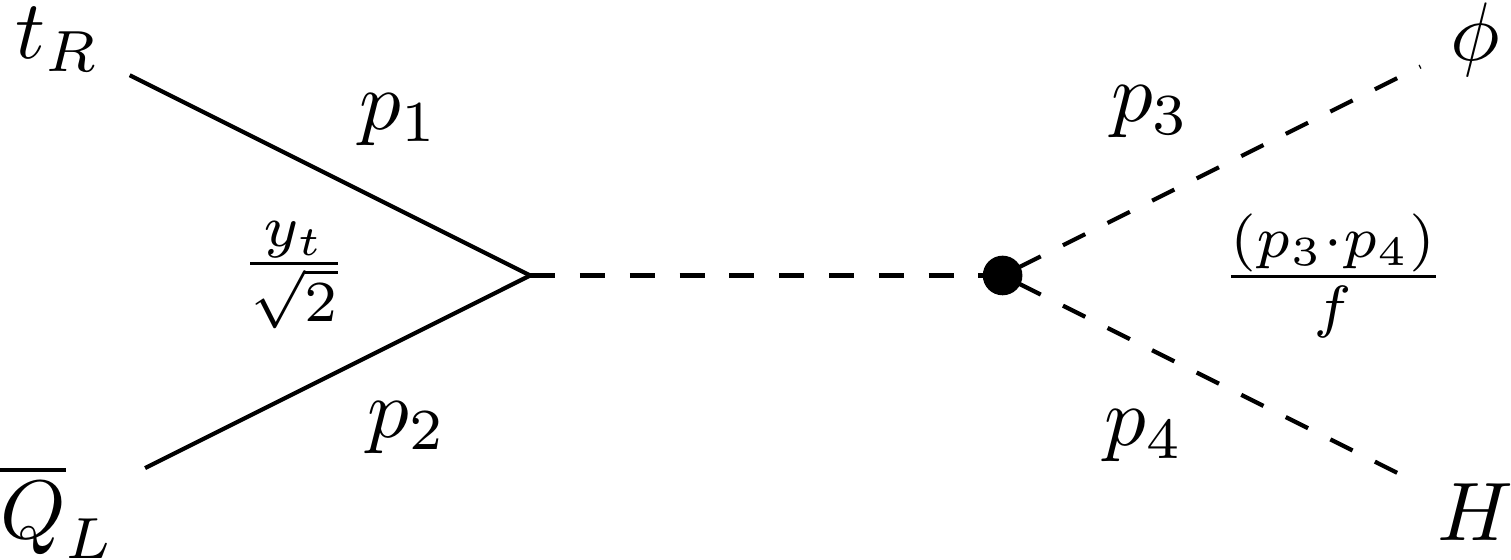}~~~
    \includegraphics[width=5.2cm]{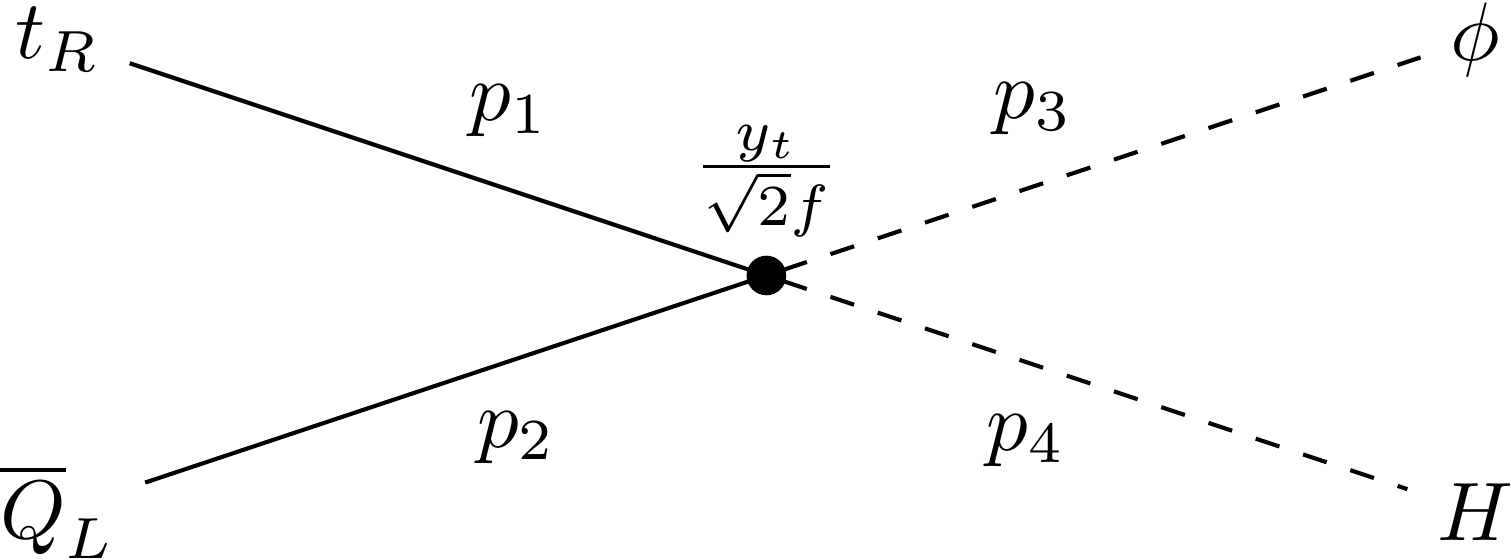}
    \caption{Single $\phi$ production channels. Note that in addition to these diagrams, $t_RH (\overline Q_L H)\to Q_L\phi (\overline t_R \phi)$ also exist. }
    \label{fig:SinglePhi}
\end{figure*}

Before discussing each case in detail, we make criteria to assess whether $\phi$ is in the thermal bath or not.
The reaction processes that bring $\phi$ into the thermal bath are the ones involving the top Yukawa coupling, shown in Fig.~\ref{fig:SinglePhi}.
The reaction rate is computed as
\beq
R_{\rm Yuk}(T)=\beta_{\rm Yuk}\frac{T^6}{f^2}
\label{eq:RYuk}
\eeq
with $\beta_{\rm Yuk}=567\zeta^2(3)y_t^2/256\pi^5\simeq0.01$~\cite{Brax:2021gpe}.
In a radiation-dominated Universe, $H \propto T^2$ and hence $R_{\rm Yuk}/n_\gamma H \propto T$ with $n_\gamma =2\zeta(3)T^3/\pi^2$ being the number density of photon is maximized at $T_{\rm reh}$.
Then, evaluating the ratio at $T = T_{\rm reh}$ as 
\bea \label{eq:cond-thermal}
\left.
\frac{R_{\rm Yuk}}{n_\gamma H}
\right|_{T=T_{\rm reh}}
=
&&0.3y_t^2 \left(\frac{106.75}{g_{\rm reh}}\right)^{1/2}
\nonumber \\
&&
\times 
\left(\frac{T_{\rm reh}}{10^{7}~{\rm GeV}}\right)
\left(\frac{10^{14}~{\rm GeV}}{f}\right)^2,
\label{eq:thermalization}
\eea
one can see whether $\phi$ had been thermalized (i.e., $\left. R_{\rm Yuk}/n_\gamma H
\right|_{T=T_{\rm reh}}>1$) or not for a given scale $f$.
Note that Eq.~(\ref{eq:cond-thermal}) is applicable for $T \gtrsim m_h$ with $m_h$ being the mass of Higgs (but $y_t$ replaced to $y_b$ if $T \lesssim m_t$).
It can also be used to find the freeze-out temperature of $\phi$ ($T_{\rm fo}$) for a given $f$ if $T_{\rm reh} > T_{\rm fo}$. 
\footnote{The ratio should be evaluated at $T_{\rm max}$ to be more precise, but we regard $T_{\rm reh}$ as a maximal temperature in the following argument, since our results are not sensitive to $T_{\rm max}$.}

Depending on its lifetime, the would-be abundance of $\phi$ is constrained by BBN \cite{Kawasaki:2017bqm} or CMB \cite{Slatyer:2016qyl} even if it never dominates the Universe.
For example, if $m_\phi \gtrsim \mathcal{O}(10) {\rm MeV}$, dissociations of Deuterium is possible.
Then, for $\tau_\phi \gtrsim 10^8 {\rm sec}$, $m_\phi Y_\phi \lesssim 10^{-14} {\rm GeV}$ is required. 
However, if $m_\phi \lesssim \mathcal{O}(1) {\rm MeV}$, BBN bound is irrelevant. 
Meanwhile, for $1 < m_\phi/{\rm MeV} \lesssim 10$, the CMB constraint on the lifetime of dark matter is $\tau_{\rm DM} \gtrsim 10^{25} {\rm sec}$ \cite{Slatyer:2016qyl}.
This bound can be interpreted as a constraint on the abundance of $\phi$ before its decay:
\beq
f_\phi \lesssim \frac{\tau_\phi}{\tau_{\rm DM}^{\rm bnd}}
\eeq
where $f_\phi$ is the would-be fractional abundance of $\phi$ relative to the observed abundance of dark matter if the mass of $\phi$ were the same as that of the decaying dark matter.

\subsection{V-A. Light-Thermal $\phi$}

\begin{center}
\begin{figure*}
    \centering
    \includegraphics[width=5.2cm]{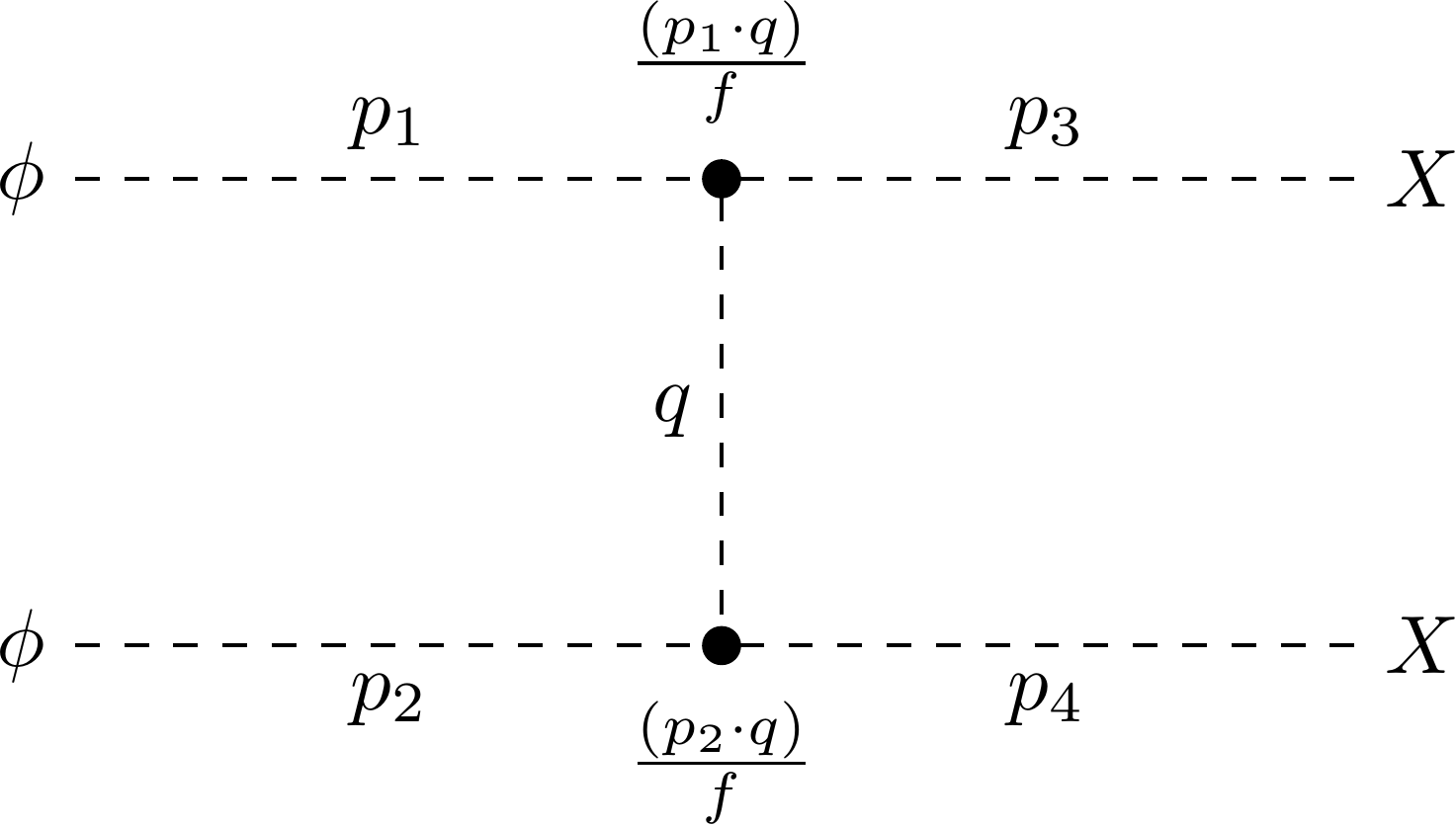}~~~
    \includegraphics[width=5.2cm]{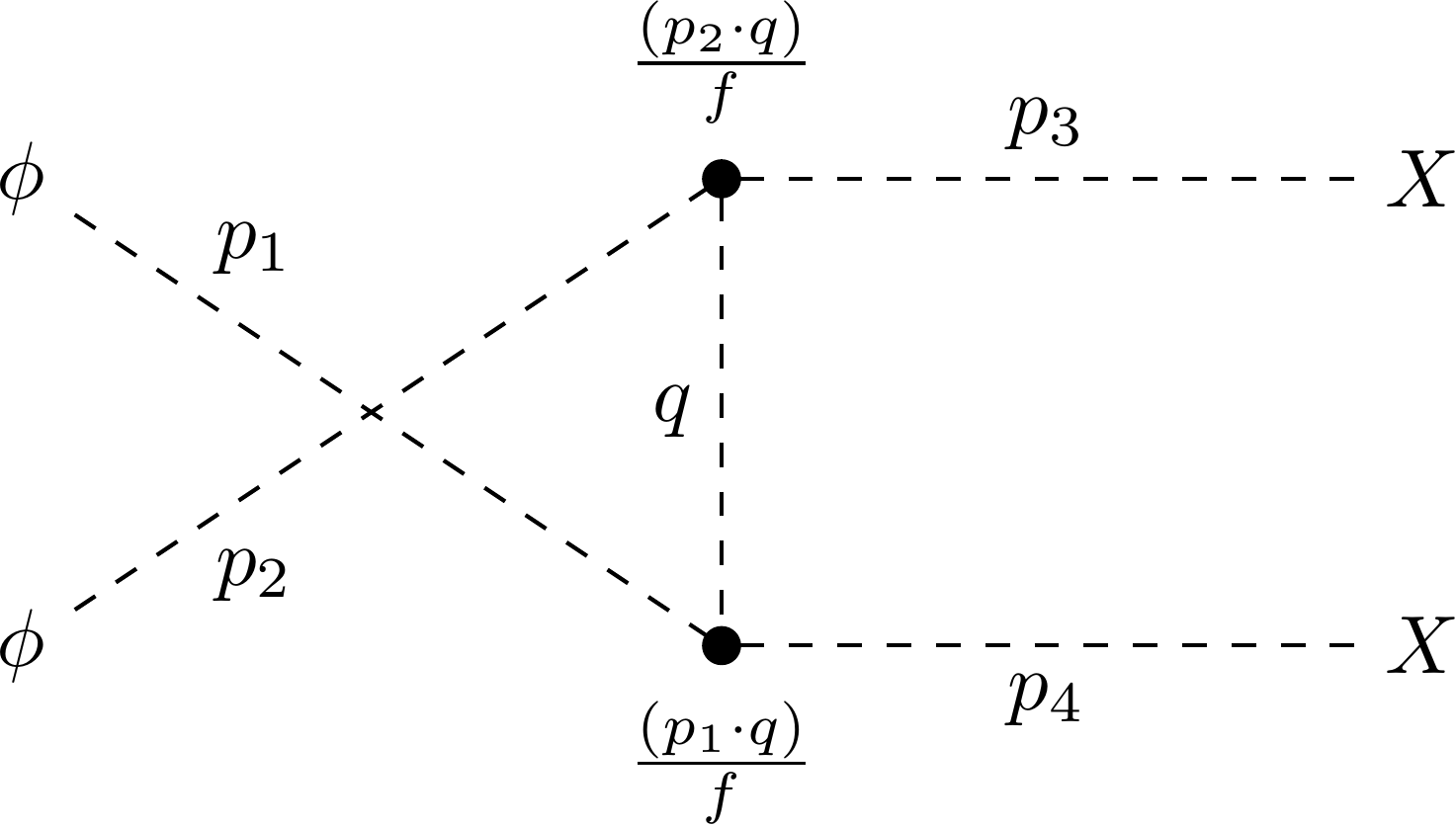}~~~
    \includegraphics[width=5.2cm]{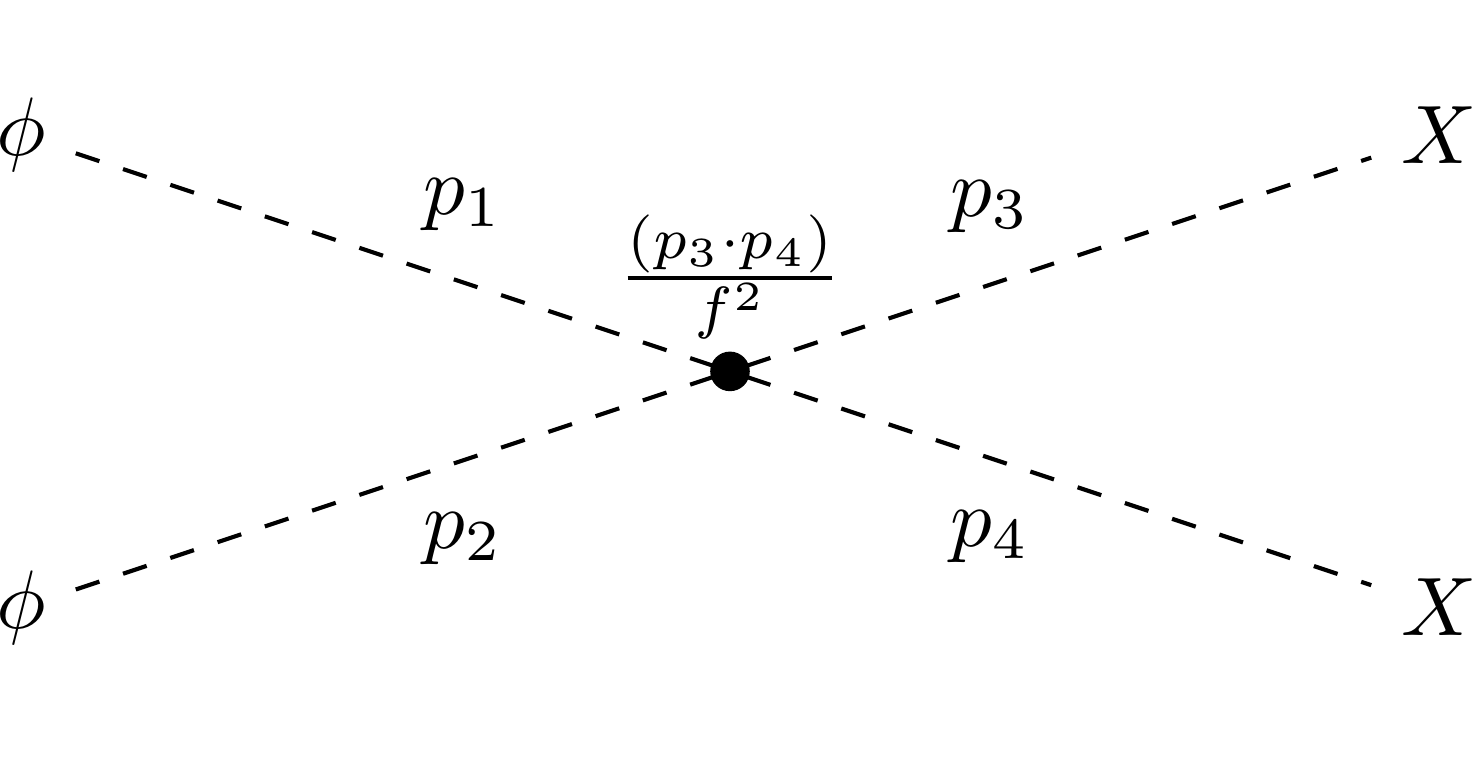}
    \caption{Feynman diagrams for the dark matter production processes when the light $\phi$ is thermal.}
    \label{fig:light-thermal}
\end{figure*}
\end{center}

When $\phi$ is lighter than the dark matter so that the decay into a pair of the dark matter particles is kinematically forbidden, the main dark matter production channel is the annihilation of either of $\phi$ or the Standard Model Higgs (through $\phi$ exchange).
In addition to the condition $m_\phi<2m_{\rm DM}$, when $\phi$ is in the thermal bath, then a pair annihilation of $\phi$ mainly produces dark matter, whose diagrams for the scalar dark matter case are shown in Fig.~\ref{fig:light-thermal}.
For the fermionic or vectorial dark matter cases, the corresponding diagrams can be obtained by replacing $X$ with $\chi$ or $V_\mu$, respectively.

The reaction rates for $\phi\phi\to XX/\chi\overline\chi$ are readily obtained as
\bea
R(T)=
\begin{cases}
\beta_0\frac{T^8}{f^4} & \text{(scalar dark matter)}\\
\beta_{1/2}\frac{m_\chi^2T^6}{f^4} & \text{(fermionic dark matter)}
\end{cases},
\label{eq:RXXFF}
\eea
where $\beta_0=\pi^3/19200\simeq 1.6\times10^{-3}$ and $\beta_{1/
2}=\zeta^2(3)/32\pi^5\simeq 1.5\times10^{-4}$.
For the vectorial dark matter case, as we see in the conformal Higgs portal scenario, the longitudinal mode is the dominant component in the dark matter number density, and thus the reaction rate becomes the same as the scalar dark matter case.
In the following, we thus consider only the scalar and fermionic dark matter cases.

From Eqs.~(\ref{eq:YDMscat}) and (\ref{eq:Oh2scat}), the dark matter relic density can be computed as
\begin{widetext}
\bea \label{eq:OhDMscat}
\frac{\Omega^{\rm scat}_{\rm DM}h^2}{0.1}\simeq
\left(\frac{106.75}{g_{\rm reh}}\right)^{3/2}
\left(\frac{10^7~{\rm GeV}}{f}\right)^4
\times
\begin{cases}
\left(\frac{T_{\rm reh}}{\rm TeV}\right)^3\left(\frac{m_X}{{150}~{\rm keV}}\right) & \text{(scalar dark matter)}\\[10pt]
\left(\frac{T_{\rm reh}}{\rm TeV}\right)\left(\frac{m_\chi}{10~{\rm GeV}}\right)^3 & \text{(fermionic dark matter)}
\end{cases},
\eea
\end{widetext}
where we have assumed $m_\phi\ll m_{X,\chi}$.
For the scale of $f$ used in Eq.~(\ref{eq:OhDMscat}), decoupling of $\phi$ from thermal bath takes place at $T \sim m_h$ thanks to bottom quark replacing top in Fig.~\ref{fig:SinglePhi}.
In this case, even if $\phi$ is lighter than ${\rm MeV}$ scale, its contribution to $\Delta N_{\rm eff}$ which is interpreted as extra neutrino contributions is negligible.
However, if $m_\phi < 2 m_e$ with $m_e$ being the mass of electron, $\phi$ can decay dominantly to photons via 1-loop diagram.
Such late decays of $\phi$ can cause an era of matter-domination during the epoch of BBN, ruining it.
In order to avoid such a danger, we assume $m_\phi > 2 m_e$ in this case of light thermal $\phi$.\footnote{Note that if $m_\phi\ll m_\nu$, such a light $\phi$ can not be constrained by BBN and/or CMB. Instead, one may consider fifth-force searches for such a long distance effect. (See, for instance, Refs.~\cite{Adelberger:2003zx,Kapner:2006si}.)}

To make sure that the dark matter stays nonthermal, we compare the reaction rate and the Hubble:
\begin{widetext}
\bea \label{eq:cond-nonthermal-dm}
\left.
\frac{R}{n_\gamma H}
\right|_{T=T_{\rm reh}}=
&&
\left(\frac{106.75}{g_{\rm reh}}\right)^{1/2}
\left(\frac{10^7~{\rm GeV}}{f}\right)^4
\times
\begin{cases}
 3.4\times10^{-4}\left(\frac{T_{\rm reh}}{\rm TeV}\right)^3\left(\frac{T_{\rm max}}{T_{\rm reh}}\right)&\text{(scalar dark matter)}\\[10pt]
8.5\times10^{-10}\left(\frac{T_{\rm reh}}{\rm TeV}\right)\left(\frac{m_\chi}{10~{\rm GeV}}\right)^2&\text{(fermionic dark matter)}
\end{cases},
\eea
\end{widetext}
from which it is evident that the dark matter abundance we obtained is consistent with the nonthermal dark matter assumption.
Notice that the scalar dark matter case depends on $T_{\rm max}$ since $R\propto T^8$ while $n_\gamma H\propto T^7$ for $T_{\rm reh}<T<T_{\rm max}$, and thus the ratio is sensitive to $T_{\rm max}$.
This is, however, not the case for the fermionic dark matter case as $R$ is suppressed by $m_\chi^2$.
Nevertheless, the dark matter is dominantly produced at $T_{\rm reh}$.
Therefore, as long as $T_{\rm reh}\gtrsim m_h$, $\phi$ is in thermal equilibrium, while the dark matter stays nonthermal.
Note that in this case a low scale inflation scenario may be preferred to avoid unitarity violation at $T_{\rm max}$ since from Eqs.~(\ref{eq:Tmax}) and (\ref{eq:Treh}) the ratio $T_{\rm max}/T_{\rm reh}$ increases for lower $T_{\rm reh}$.
A more detail on the unitarity bound in the conformal mediator scenario is discussed in Appendix.

\subsection{V-B. Light-Nonthermal $\phi$}

\begin{center}
\begin{figure*}
    \centering
    \includegraphics[width=5.2cm]{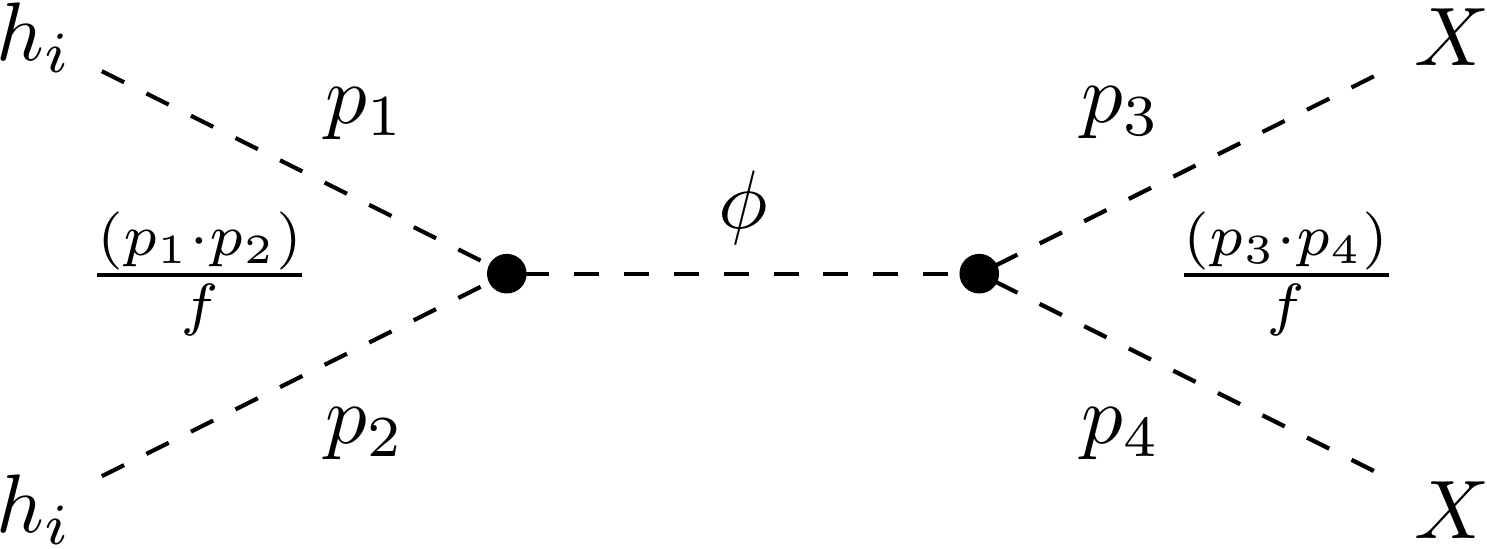}~~~
    \includegraphics[width=5.2cm]{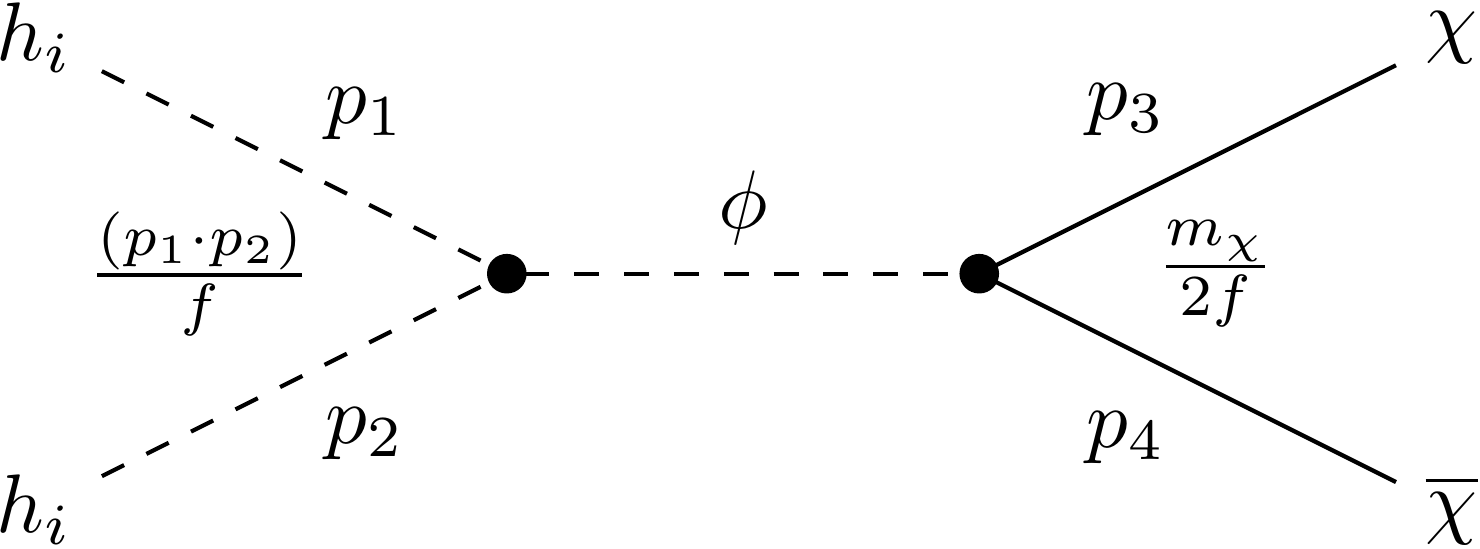}
    \caption{Feynman diagrams for the dark matter production processes for each type of dark matter when $\phi$ is nonthermal and $m_\phi<2m_{\rm DM}$.}
    \label{fig:light-nonthermal}
\end{figure*}
\end{center}

When supposing $\phi$ is light and nonthermal, the dark matter is produced via the s-channel $HH^\dagger\to\phi\to XX/\chi\overline\chi$ processes whose diagrams are shown in Fig.~\ref{fig:light-nonthermal}.

In the similar manner, we can readily compute the reaction rates whose analytic expressions are the same as Eq.~(\ref{eq:RXXFF}) but $\beta_0=\pi^3/{43200}\simeq 7.2\times10^{-4}$ and $\beta_{1/2}=\zeta^2(3)/32\pi^5\simeq1.5\times10^{-4}$.
Thus, from Eqs.~(\ref{eq:YDMscat}) and (\ref{eq:Oh2scat}) with the reaction rate for each type of dark matter, we obtain
\begin{widetext}
\beq
\frac{\Omega_{\rm DM}^{\rm scat}h^2}{0.1}\simeq
\left(\frac{106.75}{g_{\rm reh}}\right)^{3/2}\left(\frac{M_P}{f}\right)^4
\begin{cases}
\left(\frac{T_{\rm reh}}{10^{14}~{\rm GeV}}\right)^3
\times\left(\frac{m_{X}}{{1.1}~{\rm EeV}}\right) & \text{(scalar dark matter)}\\[10pt]
\left(\frac{T_{\rm reh}}{10^{14}~{\rm GeV}}\right)
\times\left(\frac{m_\chi}{{ 3.3}\times10^{12}~{\rm GeV}}\right)^3 & \text{(fermionic dark matter)}
\end{cases}.
\eeq
\end{widetext}
Conservatively speaking, even if the coherent mode is suppressed somehow, the gravitational population of $\phi$ particles at the end of inflation is expected to be $n_\phi^{\rm e} \sim \mathcal{C} H_e^3$ with $\mathcal{C} = \mathcal{O}(10^{-3})$ as long as $m_\phi \lesssim H_e$ \cite{Ema:2018ucl}.
Hence, the would-be abundance of $\phi$ before its decay well after reheating is
\bea
Y_\phi 
&\sim& \mathcal{C} \frac{T_{\rm reh} H_{\rm e}}{M_{\rm P}^2} 
\nonumber \\
&\sim& 10^{-20} \times \left( \frac{T_{\rm reh}}{10^7 {\rm GeV}} \right) \left( \frac{H_{\rm e}}{10^{13} {\rm GeV}} \right)
\eea
where we used $\mathcal{C} = 10^{-3}$ in the second line.
It can be subject to either BBN or CMB constraint.

\subsection{V-C. Heavy-Thermal $\phi$}

When $m_\phi>2m_{\rm DM}$, the dark matter is produced by the decay of $\phi$, instead of the scattering of $\phi$ or the Standard Model Higgs.
The reaction rate can be written as
\bea
R_{\rm D}(T)=\frac{m_\phi^3\Gamma_\phi {\rm Br}_{\rm DM}}{2\pi^2}\int_1^\infty\frac{\sqrt{x^2-1}}{e^{zx}-1}dx,
\label{eq:R_D}
\eea
where $z\equiv m_\phi/T$,  $\Gamma_\phi$ is the total decay width of $\phi$, and the branching ratio to a pair of the dark matter is defined as
\beq
{\rm Br}_{\rm DM}\equiv\frac{2\Gamma_{\phi\to {\rm DM}}}{\Gamma_{\phi}}
\eeq
with a factor of 2 in the numerator being the number of dark matter particles produced per decay.
For simplicity, we suppose that $\phi$ is heavier than the Standard Model Higgs, and thus the main decay channel is into a pair of the Higgs bosons:
\beq \label{eq:Gammaphi-to-HH}
\Gamma_{\phi\to HH^\dagger} \simeq \frac{m_\phi^3}{32\pi f^2}.
\eeq
Here, note that $\phi$ decays into 4 degrees of freedom of $H$ as long as the decay happens before electroweak symmetry breaking.
The decay width of $\phi$ into a pair of the scalar dark matters is estimated as $\Gamma_{\phi\to XX}\simeq(1/4)\Gamma_{\phi\to HH^\dagger}$ where a factor of 1/4 arises from the difference of the degrees of freedom of the final state.
In the vectorial dark matter case, the dominant component in the $\phi$ decay is the longitudinal mode, and thus its decay width is the same as the scalar dark matter case.
For the fermionic dark matter, the decay width is given by $\Gamma_{\phi\to \chi\overline\chi}\simeq m_\chi^2m_\phi/32\pi f^2$.

There are three scales characterizing the dark matter production: $m_\phi$, $T_{\rm d}$, and $T_{\rm fo}$ with $T_{\rm d}$ and $T_{\rm fo}$ being the temperature at which $\phi$ decays and the freeze-out temperature of $\phi$, respectively.
From Eq.~(\ref{eq:Gammaphi-to-HH}) the decay temperature can be computed as
\beq
H(T_{\rm d})=\Gamma_{\phi\to HH^\dagger} \Rightarrow T_{\rm d}=\sqrt{\frac{M_Pm_\phi^3}{32\pi\alpha_{\rm d}f^2}}.
\label{eq:Td}
\eeq
The freeze-out temperature of $\phi$ is readily obtained from Eq.~(\ref{eq:thermalization}) as
\beq
\left.\frac{R_{\rm Yuk}}{n_\gamma H}\right|_{T=T_{\rm fo}}=1\Rightarrow T_{\rm fo}=\frac{2\zeta(3)}{\pi^2}\frac{\alpha_{\rm fo}f^2}{\beta_{\rm Yuk}M_P},
\eeq
leading to
\beq
T_{\rm d}:T_{\rm fo}:m_\phi=(32\pi\alpha_{\rm d})^{1/2}x:\frac{2\zeta(3)\alpha_{\rm fo}}{\pi^2\beta_{\rm Yuk}}x^{-2}:1
\eeq
with $x\equiv\sqrt{M_Pm_\phi}/f$.
It is important to notice that for $x<(2\zeta(3)\alpha_{\rm fo}/\pi^2\beta_{\rm Yuk})^{1/2}(\simeq 9.1$ if $g_{\rm fo}=106.75$), $\phi$ decouples before its number density $n_\phi$ receives exponential suppression, namely, $T_{\rm fo}>m_\phi$.
In the same manner, the region given by $x>(2\zeta(3)\alpha_{\rm fo}/\pi^2\beta_{\rm Yuk})^{1/2}$ corresponds to $T_{\rm fo}<m_\phi$.
These two cases are distinctive for the dark matter production, in particular, for the production dominated in the IR regime.

In either case, however, there also exists the dark matter production dominated in the UV regime, namely, the scattering contribution from the thermal $\phi$ at $T_{\rm reh}$.
Such contribution is the same as given by Eq.~(\ref{eq:OhDMscat}).
As will be shown in the following, the scattering contribution becomes in general non-negligible for larger $T_{\rm reh}$ and/or smaller $f$.

When $T_{\rm fo}>m_\phi$, the dominant contribution in the IR comes from the decay of $\phi$ after the decoupling.
In this case, the decay happens at $T_{\rm d}<T_{\rm fo}$, and thus the number of produced $\phi$ is redshifted by then.
The number density of $\phi$ at $T_{\rm fo}$ is given by $n_\phi(T_{\rm fo})=(\zeta(3)/\pi^2)T_{\rm fo}^3$, and thus we obtain
\beq
Y_{\rm DM}=\frac{n_{\rm DM}(T_{\rm d})}{T_{\rm d}^3}={\rm Br}_{\rm DM}n_\phi(T_{\rm d})={\rm Br}_{\rm DM}\frac{g_{\rm d}}{g_{\rm fo}}\frac{\zeta(3)}{\pi^3}.
\eeq
Therefore, including the scattering contribution, we obtain
\begin{widetext}
\bea
\frac{\Omega_{\rm DM}h^2}{0.1}\simeq
&&
\left(\frac{106.75}{g_{\rm fo}}\right)
\begin{cases}
\left(\frac{m_X}{14~{\rm MeV}}\right)\\[10pt]
\left(\frac{\rm TeV}{m_\phi}\right)^2\left(\frac{m_\chi}{19~{\rm GeV}}\right)^3
\end{cases}
+
\left(\frac{106.75}{g_{\rm reh}}\right)^{3/2}
\left(\frac{10^{10}~{\rm GeV}}{f}\right)^4
\times
\begin{cases}
\left(\frac{T_{\rm reh}}{10^5~{\rm GeV}}\right)^3\left(\frac{m_X}{{330}~{\rm GeV}}\right)
\\[10pt]
\left(\frac{T_{\rm reh}}{10^5~{\rm GeV}}\right)\left(\frac{m_\chi}{{\rm TeV}}\right)^3
\end{cases},
\eea
\end{widetext}
where the second term corresponds to the scattering contribution, and the top and bottom lines in each term correspond to the scalar and fermionic dark matter cases, respectively.

When $T_{\rm fo}<m_\phi$, the out-of-equilibrium decay of $\phi$ does not contribute as $n_\phi(T)$ is exponentially suppressed at $T=T_{\rm fo}$.
Instead, with $T_{\rm d} > T_{\rm fo}$, the freeze-in from the on-shell $\phi$ in the thermal bath may produce the dark matter, whose reaction rate is given by Eq.~(\ref{eq:R_D}).
In particular, the dark matter is dominantly produced when $\phi$ is nonrelativistic, and thus we may use the Maxwell-Boltzmann distribution for $\phi$, instead of the Bose-Einstein distribution, resulting in $R_{\rm D}\simeq (m_\phi^2 T\Gamma_\phi {\rm Br}_{\rm DM}/2\pi^2)K_1(m_\phi/T)$ with $K_1$ being the modified Bessel function of the second kind.
The abundance of dark matter can be obtained from Eq.~(\ref{eq:BoltzmannDM3}) with $R_{\rm D}$.
Integrating the equation for $0 < T < \infty$, we obtain
\beq
Y^{\rm dec}_{\rm DM}= \frac{3}{4\pi}\frac{M_P\Gamma_\phi {\rm Br}_{\rm DM}}{\alpha(m_\phi)m_\phi^2}.
\label{eq:Ymphi}
\eeq
It is dominated by the contribution for $1 \lesssim m_\phi/T \lesssim 10$.
Hence, as long as $T_{\rm fo} \ll m_\phi$, Eq.~(\ref{eq:Ymphi}) provides a good enough approximation for the abundance of dark matter.
For the scalar dark matter, we obtain
\beq
\frac{\Omega_X^{\rm dec}h^2}{0.1}\simeq \left(\frac{106.75}{g_*(m_\phi)}\right)^2x^4\left(\frac{m_X}{3.3~{\rm keV}}\right),
\eeq
where by noticing that $x>9.1$ (assuming $g_*(m_\phi)=106.75$) and $m_X\lesssim10$ keV is excluded by the Lyman-$\alpha$ bound~\cite{Narayanan:2000tp,Viel:2005qj,Viel:2013fqw,Baur:2015jsy,Irsic:2017ixq,Palanque-Delabrouille:2019iyz,Garzilli:2019qki,Ballesteros:2020adh}, we may conclude that the scalar dark matter scenario is not viable in this parameter space due to the overproduction of the dark matter.
For the fermionic dark matter, we obtain
\beq
\frac{\Omega^{\rm dec}_\chi h^2}{0.1}\simeq 
\left(\frac{106.75}{g_*(m_\phi)}\right)^2\left(\frac{x}{10}\right)^4\left(\frac{\rm TeV}{m_\phi}\right)^2\left(\frac{m_\chi}{44~{\rm MeV}}\right)^3.
\eeq
Note that in addition to the decay contribution, there also exists the scattering contribution as was the case when $T_{\rm fo}>m_\phi$. 

\subsection{V-D. Heavy-Nonthermal $\phi$}

When $\phi$ is nonthermal, it is dominantly produced through the Yukawa interactions, whose reaction rate is given by Eq.~(\ref{eq:RYuk}).
Therefore, the yield value of $\phi$ is readily computed from Eq. (\ref{eq:YDMscat}) as
\beq
Y^{\rm Yuk}_\phi = \frac{\beta_{\rm Yuk} M_P T_{\rm reh}}{\alpha_{\rm reh}f^2}.
\label{eq:Yphi}
\eeq

In addition to the Yukawa interaction contributions, the inverse decay process, $HH^\dagger\to\phi$, may produce a single $\phi$ as well.
Notice that such a process is possible only when $\phi$ is heavier than the Higgs boson.
When $\phi$ is sufficiently heavier than the Standard Model Higgs, one may take the massless limit for the Higgs.
Under this setup, the reaction rate is readily obtained from Eq.~(\ref{eq:ReactionRate}) as
\beq
R_{\rm ID}(T)\simeq \frac{m_\phi^6}{32\pi^3f^2}I(z),
\eeq
where $I(z)$ is defined as
\bea
I(z) =&&
\int_1^\infty \diff x_+ \int_{\sqrt{x_+^2-1}}^{\sqrt{x_+^2-1}}\diff x_-\nonumber
\\&&
\times\left\{
\frac{1}{e^{\frac{z}{2}(x_++x_-)}-1}\frac{1}{e^{\frac{z}{2}(x_+-x_-)}-1}
\right\}.
\eea
To a good approximation, we may use
\bea
I(z)\simeq K_{5/2}(z)\simeq
\begin{cases}
3\sqrt{\frac{\pi}{2}}\frac{1}{z^{5/2}}\qquad (z\ll 1)\\[10pt]
\sqrt{\frac{\pi}{2}}\frac{1}{z^{1/2}}e^{-z}\qquad (z\gtrsim 1)
\end{cases},
\eea
where $K_{5/2}(z)$ is the modified Bessel function of the second kind.
Therefore, for $T\gg m_\phi$, we obtain
\beq
R_{\rm ID}(T)\simeq \frac{3m_\phi^{7/2}T^{5/2}}{32\sqrt{2}\pi^{5/2}f^2},
\eeq
resulting in
\beq
Y^{\rm ID}_\phi\simeq \frac{3M_Pm_\phi}{80\sqrt{2}\alpha(m_\phi)\pi^{5/2}f^2}.
\eeq
On the other hand, by comparing $Y^{\rm Yuk}_\phi$ and $Y^{\rm ID}_\phi$, we find
\beq
\frac{Y^{\rm ID}_\phi}{Y^{\rm Yuk}_\phi}\simeq 0.02\left(\frac{g_{\rm reh}}{g_*(m_\phi)}\right)^{3/2}\left(\frac{m_\phi}{T_{\rm reh}}\right),
\eeq
and thus, the inverse decay contribution turns out to be negligible in our case.
Therefore, the resultant dark matter relic abundance becomes
\bea
\frac{\Omega^{\rm dec}_{\rm DM}h^2}{0.1}\simeq
&&
\left(\frac{106.75}{g_{\rm reh}}\right)^{3/2}\left(\frac{M_P}{f}\right)
\nonumber\\&&
\times\left(\frac{T_{\rm reh}}{10^{10}~{\rm GeV}}\right)\left(\frac{{\rm Br}_{\rm DM}m_{\rm DM}}{140~{\rm GeV}}\right).
\eea

\section{VI. Phenomenology}

When the mediator is lighter than the dark matter, direct detection experiments may have sensitivity to the parameter spaces with relatively low reheating temperature.
The direct detection constraint on the dark matter-nucleon scattering cross section usually assumes an effective contact interaction.
On the other hand, if the mediator mass is sufficiently light, such as $m_\phi\lesssim 40$ MeV, the nuclear recoil through the light mediator exchange can mimic the contact interaction event, which allows us to look for the FIMP dark matter through the light mediator \cite{Fornengo:2011sz,Hambye:2018dpi}.
Our analysis closely follows Ref.~\cite{Hambye:2018dpi}.
The relevant formulas and parameters are summarized in Appendix.

The strategy to give a constraint by the XENON1T data \cite{Aprile:2018dbl} is the following.
We use the experimental constraint on the dark matter-nucleon cross section $\sigma_{\rm DM-n}$ to evaluate the differential recoil rate given in Eq.~(\ref{eq:DRecoil}).
For the dark matter-nucleus differential cross section in our case, we obtain
\bea
b(q)=\frac{m_Nm_{\rm DM}^2[Zm_p+(A-Z)m_n]^2}{32\pi f^4(q^2+m_\phi^2)^2}F^2(q),
\label{eq:b_theory}
\eea
where $b(q)$ is defined in Eq.~(\ref{eq:defb}) with $q=\sqrt{2m_NE_R}$, and $m_p$ and $m_n$ are the proton and neutron masses, respectively.
Note that for both scalar and fermionic dark matter cases the expression for $b(q)$ becomes the same. 
With this we can compute $\diff R/\diff E_R$ for both experimental input and theoretical prediction.
Using the efficiency factor $\epsilon(E_R)$ taken from Ref.~\cite{Aprile:2018dbl}, we minimize
\bea
\Delta^2_{\rm DR}\equiv\frac{1}{R_{\rm exp}^2}\int \diff E_R \epsilon^2(E_R)\left[\frac{\diff R_{\rm exp}}{\diff E_R}-\frac{\diff R_{\rm th}}{\diff E_R}\right]^2
\eea
by taking $m_{\rm DM}$ and $f$ as free parameters, where $R_{\rm exp}$ and $R_{\rm th}$ are the recoil rates evaluated by XENON1T data and by Eqs.~(\ref{eq:b_theory}) and (\ref{eq:DRecoil}), respectively.

\begin{figure*}[th]
    \centering
    \includegraphics[width=15cm]{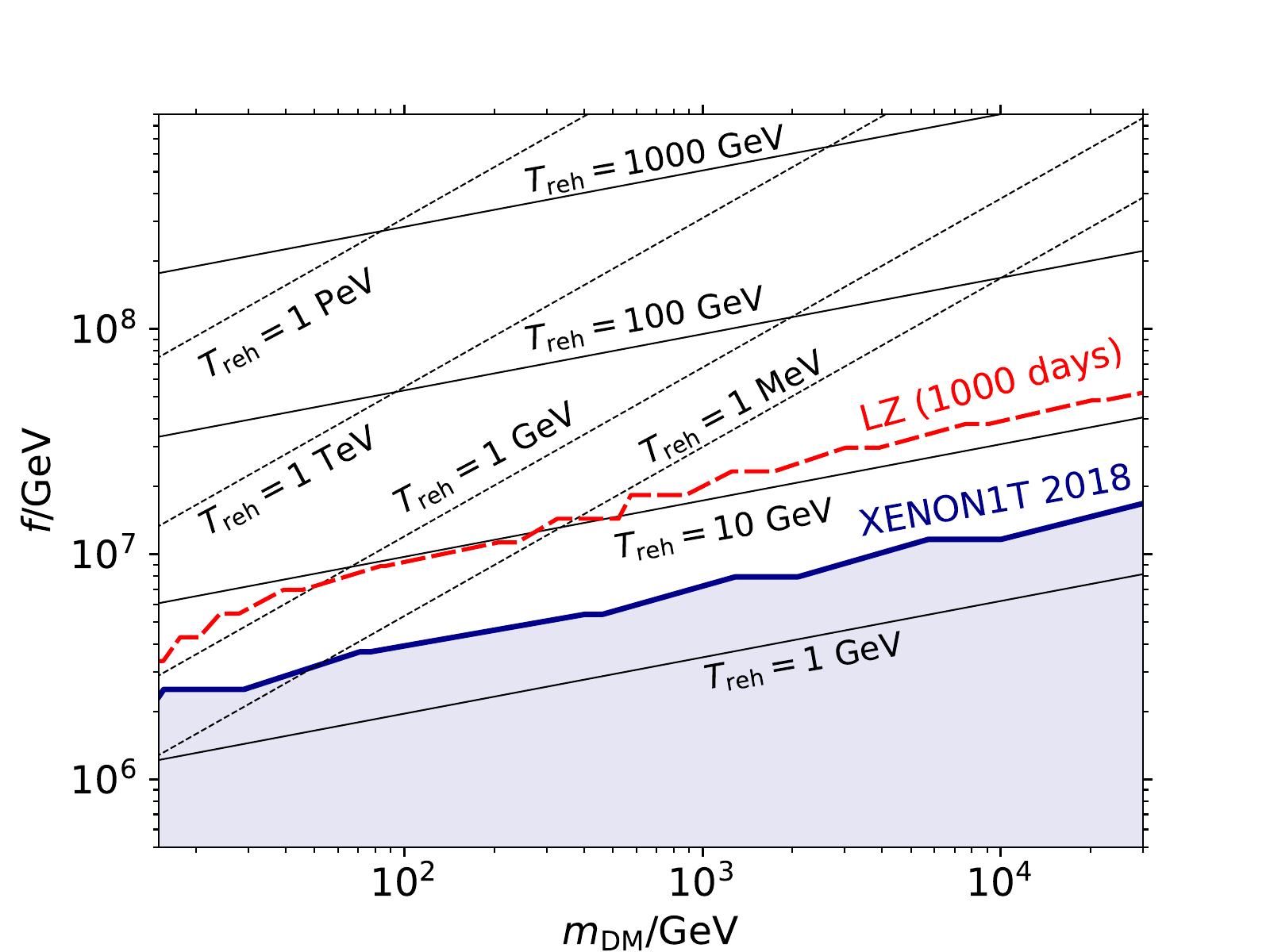}
    \caption{Constraints from XENON1T. The black solid lines correspond to the scalar dark matter with the light-thermal mediator, while the black dashed lines  show the fermionic dark matter case with the light-thermal mediator. The red long-dashed line is the projection of the LZ experiment with 1000 days of exposure.
    }
    \label{fig:DD}
\end{figure*}

Figure \ref{fig:DD} shows the result of the direct detection constraint given by using XENON1T data.
The black solid lines depict the correct dark matter relic density for the scenario with the scalar dark matter and the thermal $\phi$, where we take $m_\phi=0$ to a good approximation up to $m_\phi\lesssim 40$ MeV.
Note that our analysis does not apply for a heavier $\phi$, since the recoil energy distribution is largely deformed by the effect of the nonzero mass of $\phi$, and the direct detection experiments start losing the sensitivity because of the absence of the enhancement in the cross section from the light mediator.
The black dashed lines are the case of the fermionic dark matter with fixed $T_{\rm reh}$, while the black solid lines correspond to the scalar dark matter case.
Note that the vectorial dark matter is the same as the scalar dark matter in the figure.
The LZ projection with 1000 days exposure is shown by the red dense-dashed line, where the data is taken from Ref.~\cite{Akerib:2018lyp}.
Notice that the direct detection constraints on $f$ scale as $m_{\rm DM}^{1/4}$ since the recoil rate is proportional to $m_{\rm DM}/f^4$.

Before concluding the section, we consider possible cosmological limits on such a light $\phi$.
As mentioned above, our direct detection constraint applies only when $m_\phi\lesssim40$ MeV with which a dominant decay channel is into a pair of electrons.
By using the decay width $\Gamma_{\phi\to ee}\simeq m_e^2m_\phi/(32\pi f^2)$, the decay temperature is obtained as
\beq
T_{\rm d}=\sqrt{\frac{\Gamma_{\phi\to ee}M_P}{\alpha_{\rm d}}}.
\eeq
On the other hand, if $T_{\rm fo} \gtrsim m_\phi$, the energy density of the produced $\phi$ dominates over the radiation at the temperature $T_{\rm dom}=30\zeta(3)m_\phi/g_{\rm fo}\pi^4$.
Requiring that $\phi$ should decay before it dominates in the energy density, we obtain a limit given by
\bea
T_{\rm dom}<T_{\rm d} \Rightarrow f&<&3.8\times10^8~{\rm GeV}\left(\frac{g_{\rm fo}}{106.75}\right)
\nonumber\\&&
\times\left(\frac{10.75}{g_{\rm d}}\right)^{1/2}\left(\frac{10~{\rm MeV}}{m_\phi}\right)^{1/2}.
\eea
Furthermore, $T_{\rm d}$ should also be higher than the BBN temperature ($T_{\rm BBN}\sim 1$ MeV), which leads to a constraint given by
\beq
f\lesssim 10^7~{\rm GeV} \left(\frac{10.75}{g_{\rm d}}\right)^{1/2}\left(\frac{m_\phi}{10~{\rm MeV}}\right)^{1/2}.
\eeq
Therefore, the region of $f\gtrsim 10^7$ GeV in Fig.~\ref{fig:DD} may conflict with these constraints which can, however, be easily avoided if $m_\phi$ is heavier.
Nevertheless, in such a case, our direct detection constraint needs to be replaced by the one based on a contact interaction approximation between dark matter and the nucleus.

\section{VII. Conclusion}

In this work, a new portal coupling to dark matter is discussed, where the portal sector is introduced through a conformal factor of the space-time metric.
The introduction of such conformal factor is equivalent to considering a scalar field that couples to the Ricci scalar in the Jordan frame, and consequently the scalar particle may couple to the trace of the energy-momentum tensor of the matter sector, including the dark matter sector.
Supposing that the dark matter sector is secluded from the Standard Model sector, we consider two examples of choosing the conformal factor: conformal Higgs portal and conformally induced mediator portal.

In the conformal Higgs portal scenario, we show that the well-known Higgs nonminimal coupling can produce the secluded dark matter via freeze-in.
We emphasize that this scenario is consistent with the Higgs inflation, and a sharp prediction can be obtained on the dark matter mass: 71 keV for the scalar (vector) dark matter and 1.7 PeV for the fermionic dark matter.

When the scalar field, $\phi$, induced through the conformal factor plays a role of a mediator propagating between the dark matter sector and the Standard Model, new dark matter production channels are allowed, depending on whether or not $\phi$ is in the thermal bath, and whether or not the decay of $\phi$ into a pair of dark matter is kinematically allowed.
In particular, we show that when $\phi$ is light and thermalized, the dark matter direct detection experiments can test the model.
The CMB and BBN constraints on the late time decay of $\phi$ are also discussed.
\newline

{\bf Acknowledgements:}
The work was supported by a KIAS Individual Grant No. PG080301 (KK) and No. PG021403 (PK) at Korea Institute for Advanced Study, by National Research Foundation of Korea (NRF) Grant No. NRF-2019R1A2C3005009 (PK), funded by the Korea government (MSIT), by Research Base Construction Fund Support Program (WIP) funded by Jeonbuk National University in 2021, and by Basic Science Research Program through the National Research Foundation of Korea (NRF) funded by the Ministry of Education (No. 2017R1D1A1B06035959) (WIP).

\appendix

\section{APPENDIX}

\subsection{A. Conformally Introduced Scalar Sector}

We here give an explicit correspondence between the conformally coupled scalar sector in the Jordan frame and that in the Einstein frame.
In the Jordan frame, the action of the gravity and matter sectors is given by
\bea
&&
S=S_{\rm grav}+S_{\rm matt},
\\
&&
S_{\rm grav}=\frac{M_P^2}{2}\int \diff^4x\sqrt{-\tilde g}C(\phi)\tilde R,
\\
&&
S_{\rm matt}=\int\diff^4x\sqrt{-\tilde g}{\cal L}_{\rm matt},
\eea
where we defined $\tilde g\equiv {\rm det}(\tilde g_{\mu\nu})$
and $\tilde R$ as the determinant of the metric and the Ricci scalar in the Jordan frame, respectively.
The metric defined in the Jordan frame, $\tilde g_{\mu\nu}$, and that in the Einstein frame, $g_{\mu\nu}$, are related with $\tilde g_{\mu\nu}=C^{-1}g_{\mu\nu}$, called the Weyl rescaling, such that $S_{\rm grav}$ becomes the Einstein-Hilbert action.
Indeed, under the rescaling, the Ricci scalar transforms as
\bea
C^{-1}\tilde R=
&&
R+\frac{(D-1)(D-2)}{4}g^{\mu\nu}(\nabla_\mu\ln C)(\nabla_\nu\ln C)\nonumber
\\
&&-(D-1)g^{\mu\nu}\nabla_\mu\nabla_\nu\ln C
\eea
in $D$ dimensional space time\footnote{See, for instance, Appendix D of \cite{Wald:1984rg}}, and thus we obtain
\bea
S_{\rm grav}=\frac{M_P^2}{2}\int\diff^4x\sqrt{-g}\left[R+\frac{3}{2}g^{\mu\nu}(\del_\mu\ln C)(\del_\nu\ln C)\right],\nonumber\\
\eea
where the connection terms are omitted.
Notice that the derivative term with respect to $\ln C$ appears, which may play as a canonically normalized kinetic term for $\phi$, when $C(\phi)=\exp\pm\sqrt{\frac{2}{3}}\frac{\phi}{M_P}$.
In this case therefore a kinetic term of $\phi$ in ${\cal L}_{\rm matt}$ is not needed.

The relation between $\phi$ and the canonically normalized field $\chi$ depends on the choice of $C(\phi)$ and the kinetic term in ${\cal L}_{\rm matt}$.
For instance, we may define the kinetic term in ${\cal L}_{\rm matt}$ by
\bea
{\cal L}_{\rm matt} \supset M_P^2 K\tilde g^{\mu\nu}(\del_\mu C^{1/2})(\del_\nu C^{1/2}),
\eea
leading to the kinetic term in the Einstein frame given by
\bea
S_{\phi,{\rm kin}}=\frac{3+K}{4}M_P^2\int\diff^4x\sqrt{-g}g^{\mu\nu}(\del_\mu\ln C)(\del_\nu\ln C),\nonumber\\
\eea
where $K$ is an arbitrary constant and chosen so that the kinetic term for $\phi$ is canonically normalized.
This choice gives $\chi=\phi$ when $C(\phi)=\exp\pm\alpha \phi/M_P$ with $\alpha$ being an arbitrary constant, and taking $K=2/\alpha^2-3$.

Another example is the case where the kinetic term is given by
\beq
{\cal L}_{\rm matt}\supset \frac{1}{2} \tilde g^{\mu\nu}(\del_\mu\phi)(\del_\nu\phi)
\eeq
in the Jordan frame, leading to
\beq
S_{\phi,{\rm kin}}=\int\diff^4x\sqrt{-g}\frac{1}{2}\frac{1}{(1-\xi\phi^2/M_P^2)^2}g^{\mu\nu}(\del_\mu\phi)(\del_\nu\phi),
\eeq
where we have taken $C(\phi)=1-\xi\phi^2/M_P^2$ and $\xi=1/6$, and thus we obtain
\bea
S_{\phi,{\rm kin}}=\int\diff^4x\sqrt{-g}\frac{1}{2}g^{\mu\nu}(\del_\mu\chi)(\del_\nu\chi),
\eea
where $\phi/M_P\equiv\sqrt{6}\tanh(\chi/\sqrt{6}M_P)$.

In any cases of the choice of the kinetic term, the potential of $\phi$ in the Einstein frame, $V$, is related to that in the Jordan frame, $\tilde V$, as $V=\tilde{V}/C^2$ which we implicitly take $V=(1/2)m_\phi^2\phi^2$ in the conformally induced mediator scenario in the text.
Moreover, the field displacement of $\phi$ from the origin is assumed to be zero during/at the end of inflation, so that the coherent oscillation of $\phi$ does not come into play in our discussion of dark matter production \footnote{There can be higher order self-interaction terms of $\phi$, which lifts up $V$ so as to make misalignment of $\phi$ small enough.}
.

To see the interactions among $\phi$ and the Standard Model particles, it is convenient to consider the expansion $\tilde g_{\mu\nu}\simeq g_{\mu\nu}+\delta g_{\mu\nu}$ with $\tilde g_{\mu\nu}=C^{-1}g_{\mu\nu}\simeq(1-\delta C)g_{\mu\nu}\Rightarrow\delta g_{\mu\nu}=-\delta C(\phi) g_{\mu\nu}$ where we assume $C(\phi)$ can be expanded as $C(\phi)\simeq 1+\delta C(\phi)$.
Thus, $S_{\rm matt}$ can be expanded as
\bea
&&
S_{\rm matt}\simeq\int\diff^4x\sqrt{-g}{\cal L}_{\rm matt}+\delta S_{\rm matt},\\
&&
\delta S_{\rm matt}=\frac{1}{2}\int\diff^4x\sqrt{-g}T^{\rm matt}_{\mu\nu}\delta g^{\mu\nu},
\eea
where we defined
\beq
T^{\rm matt}_{\mu\nu}=\frac{2}{\sqrt{-g}}\frac{\delta(\sqrt{-g}{\cal L}_{\rm matt})}{\delta g^{\mu\nu}}.
\eeq
From $\tilde g^{\mu\nu}\simeq g^{\mu\nu}-\delta g^{\mu\nu}$ and $\delta g^{\mu\nu}=\delta C(\phi)g^{\mu\nu}$, we obtain
\beq
\delta S_{\rm matt} = \frac{1}{2}\int\diff^4x\sqrt{-g}\delta C(\phi)g^{\mu\nu}T^{\rm matt}_{\mu\nu}.
\eeq

\subsection{B. Perturbative Unitarity}

Since our interest is in the reactions at high temperatures, the bound from perturbative unitarity is one of the concerns.
We here restrict ourselves to consider the tree-level unitarity, which is given by
\beq
|{\rm Re}~a_J (s) | < \frac{1}{2},
\eeq
where $a_J$ is the partial-wave amplitude for the total angular momentum $J$, and a Mandelstam variable $s$ is the total energy$^2$ in the CM frame.  
$a_J (s)$ is related to the tree-level scattering amplitude ${\cal M}$ through
\bea
a_J (s) = \frac{1}{32\pi}\int_{-1}^{1}{\cal M} (s,\cos\theta) P_J ( \cos\theta ) 
\diff ( \cos\theta ) ,
\eea
where $P_J (\cos\theta)$ is the Legendre polynomial of degree $J$, and $\theta$ is the polar angle of the final particle reltive to the direction of initial particles in the CM frame.
We will consider ${\cal M}$ for $2\to2$ processes with a scattering angle $\theta$ in each scenario of the conformal Higgs portal and the conformal mediator portal.

It is convenient to give a generic formula for the tree-level unitarity bound when the scattering amplitude can be parametrized by
\beq
{\cal M_{\rm S}}=\frac{s^{k/2}}{\Lambda^k}
\label{eq:M_S}
\eeq
or
\beq
{\cal M_{\rm F}}=\frac{s^{k/2}}{\Lambda^{k/2+1}}\bar v(p_2) u(p_1),
\label{eq:M_F}
\eeq
where $k$ is an integer, $\Lambda$ is a cutoff scale, and the subscripts S and F indicate scalars and fermions in the final state.
For the scalar case, we readily obtain the bound
\beq
\sqrt{s}<(8\pi)^{1/k}\Lambda.
\eeq
For the fermion case, the only helicity amplitudes with the same helicity of the two fermions can remain nonzero, and for each amplitude, by using $\bar v(p_2) u(p_1)=\sqrt{s}$, we obtain
\beq
\sqrt{s}<(8\pi)^{1/(k+1)}\Lambda.
\eeq
Since the reaction rates of our interest with the collision energy higher than $T_{\rm max}$ is exponentially suppressed and becomes irrelevant due to the thermal distribution function, we may take $\sqrt{s}\simeq T_{\rm max}$ when discussing the unitarity limit.
Unitarity limits in each case of dark matter and portal scenarios are summarized in Tab.~\ref{tab:Unitarity}.

\begin{table*}[ht]
\begin{tabular}{l|c|c|c|c|c}
    scenario & dark matter & relevant channel & ~~~$k$~~~ & $\Lambda$ & unitarity bound \\\hline
    Higgs portal & $X (V_\mu)$ & $HH^\dagger\to XX$ & $1$ & $(2/\xi)^{1/2}M_P$ & $(16\pi/\xi)^{1/2}M_P$\\
    & $\chi$ & $HH^\dagger\to \chi\bar\chi$ & $0$ & $2M_P^2/\xi m_\chi$ & $(16\pi/\xi)M_P^2/m_\chi$\\
    light/heavy-thermal $\phi$ & $X (V_\mu)$ & $\phi\phi\to XX$ & $1$ & $(2/\sqrt{3})f$ & $2(8\pi/3)^{1/2}f$\\
    & $\chi$ & $\phi\phi\to \chi\bar\chi$ & $0$ & $ 2f^2/m_\chi$ & $16\pi f^2/m_\chi$\\
    light-nonthermal $\phi$ & $X (V_\mu)$ & $HH^\dagger\to XX$ & $1$ & $2f$ & $2(8\pi)^{1/2}f$ \\
    & $\chi$ & $HH^\dagger\to \chi\bar\chi$ & $0$ & $4f^2/m_\chi$ & $32\pi f^2/m_\chi$
\end{tabular}
\caption{Summary of unitarity bounds. $k$ and $\Lambda$ are the parameters appeared in Eqs.~(\ref{eq:M_S}) and (\ref{eq:M_F}).}
\label{tab:Unitarity}
\end{table*}

\subsection{C. Basic Formulas for Direct Detection Bounds}

We summarize formulas and relevant parameters used to give the constraints by the direct detection experimental data.
For a more comprehensive review, see, for instance, Refs.~\cite{Lewin:1995rx,Schumann:2019eaa,Lin:2019uvt}.

Direct detection experiments look for the recoil events of dark matter scattering off a target material, whose recoil rate $R$ is defined per unit target mass and per unit time, and the differential recoil rate is defined by
\beq
\frac{\diff R}{\diff E_R}=N_T\int\frac{\diff \sigma_{\rm DM-N}}{\diff E_R} v\diff n_{\rm DM},
\eeq
where $E_R$ is the nucleus recoil energy, $N_T=N_A/A$ with the Avogadro number $N_A=6.02\times10^{23}$/g is the number of a nucleus of atom mass $A$ in 1 g of substance, $\sigma_{\rm DM-N}$ is the dark matter-nucleus scattering cross section, and $v$ is the dark matter velocity in the lab frame.

For the local dark matter flux, we use
\bea
v\diff n_{\rm DM}=\frac{\rho_{\rm DM}^0}{m_{\rm DM}}vf(\vec v)\diff^3\vec v,
\eea
where the local dark matter energy density $\rho_{\rm DM}^0=0.3$ GeV/cm$^3$, and $f(\vec v)$ is assumed to be Maxwellian distribution:
\beq
f(\vec v)=N^{-1}e^{-|\vec v|^2/v_0^2}\theta(v_{\rm esc}-|\vec v|)
\eeq
with $v_0$ and $v_{\rm esc}$ being the dark matter velocity and the escape velocity above which the dark matter fly away from the galaxy, respectively.
In our analysis, we take $v_0=220$ km/s and $v_{\rm esc}=550$ km/s.
To normalize as $\int f(\vec v)\diff^3 \vec v=1$, we may choose the normalization factor
\bea
N=
% &&\int_0^{v_{\rm esc}}(4\pi v^2)e^{-v^2/v_0^2}\diff v
% \nonumber\\
% =&&
\pi^{3/2}v_0^3\left[{\rm erf}\left(\frac{v_{\rm esc}}{v_0}\right)-\frac{2}{\sqrt v_0}\frac{v_{\rm esc}}{v_0}e^{-v_{\rm esc}^2/v_0^2}\right].
\eea
Taking into account the velocity of the solar system, $\vec v\to \vec v+\vec v_S$ with $|\vec v_S|=240$ km/s, we define
\bea
g(v_{\rm min})&&
\equiv \int v^{-1}f(\vec v+\vec v_S)\diff^3\vec v
\nonumber\\
&&
=\frac{\pi^{3/2}v_0^3}{2|\vec v_S|N}\left[{\rm erf}\left(\frac{v_{\rm min}+|\vec v_S|}{v_0}\right)-{\rm erf}\left(\frac{v_{\rm min}-|\vec v_S|}{v_0}\right)\right],
\nonumber\\
\eea
where $v_{\rm min}\equiv(m_N E_R/2\mu_N^2)^{1/2}$ with $m_N$ and $\mu_N$ being the nucleus mass and the reduced mass of the dark matter and nucleus, respectively.

The differential cross section for the dark matter-nucleus scattering in the lab frame is given by
\beq
\frac{\diff \sigma_{\rm DM-N}}{\diff E_R}=\frac{\overline{|{\cal M}|^2}}{32\pi m_{\rm DM}^2m_N v^2}| F(q) |^2,
\eeq
where $\overline{|{\cal M}|^2}$ is the spin-averaged squared amplitude, and $F(q)$ with $q=\sqrt{2m_N E_R}$ is the nuclear form factor which we take \cite{Helm:1956zz}
\bea
F(q)=3\frac{j_1(qr_N)}{qr_N}e^{-(qs)^2/2}
\eea
with $j_1$ being the spherical Bessel function of the first kind, and to a good approximation $r_N^2=c^2+(7/3)\pi^2 a^2-5s^2$, $c\simeq 1.23A^{1/3}-0.6$ fm, $a\simeq 0.52$ fm, and $s\simeq0.9$ fm \cite{Lewin:1995rx}.
The direct detection experiments, such as XENON1T~\cite{Aprile:2018dbl}, put a bound on the dark matter-nucleon scattering cross section $\sigma_{\rm DM-n}$ which is related to $\sigma_{\rm DM-N}$ by
\beq
\frac{\diff\sigma_{\rm DM-N}}{\diff E_R}=\frac{m_N}{2\mu_n^2}\frac{A^2}{v^2}\sigma_{\rm DM-n}
| F(q)|^2,
\eeq
where $\mu_n$ is the reduced mass of the dark matter and nucleon.
To simplify the expression, we define
\beq
\frac{\diff\sigma_{\rm DM-N}}{\diff E_R}\equiv \frac{b(E_R)}{v^2},
\label{eq:defb}
\eeq
which allows to write
\beq
\frac{\diff R}{\diff E_R}=N_T\frac{\rho_{\rm DM}^0}{m_{\rm DM}}b(E_R)g(v_{\rm min}).
\label{eq:DRecoil}
\eeq

\bibliography{biblio.bib}
\end{document}